\documentclass[twoside]{article}
\usepackage[latin1]{inputenc}
\usepackage{t1enc}
\usepackage{epsfig}
\usepackage{a4}
\usepackage{tabularx}
\usepackage{epsf}
\newcommand{\mdot}{\ensuremath{\dot{M}}}   

\def\bref{\vspace{4pt}\noindent\hangindent=10mm}

\begin{document}

\setcounter{figure}{0}
\setcounter{section}{0}
\setcounter{equation}{0}

\begin{center}
{\Large\bf
Evolution of Massive Stars along the Cosmic History}\\[0.2cm]
%Massive Stars in the Cosmos}\\[0.7cm]
{\bf
Georges Meynet, Sylvia Ekstr\"om, Cyril Georgy, Cristina Chiappini and Andr\'e Maeder \\[0.17cm]
Astronomical Observatory of the Geneva University\\
CH-1290 Versoix, Switzerland\\
georges.meynet@unige.ch
}
\end{center}

\vspace{0.5cm}

\begin{abstract}
\noindent{\it
Massive stars are ``cosmic engines'' (cf the title of the IAU Symposium 250). They drive the photometric and chemical evolution of galaxies, inject energy and momentum through stellar winds and supernova explosions, they modify in this way the
physical state of the interstellar gas and have an impact on star formation. 
The evolution of massive stars depends sensitively on the metallicity which has an impact on the
intensity of the line driven stellar winds and on rotational mixing. We can distinguish four
metallicity regimes: 1.- the Pop III regime $0 \le Z < \sim 10^{-10}$; 2.-  The low metallicity regime $10^{-10} \le Z < 0.001$;  3.-  The near solar metallicity regime $0.001 \le Z < 0.020$; 4.-  The high metallicity regime $0.020 \le Z$. In each of these metallicity ranges, some specific physical processes occur. In this review we shall discuss these physical processes and their 
consequences for nucleosynthesis and the massive star populations in galaxies. We shall mainly focus on the effects of axial rotation and mass loss by line driven winds, although of course other processes like binarity, magnetic fields, transport processes by internal waves may also play important roles.}
\end{abstract}

\section{The evolution of the metallicity during the cosmic history}

\begin{figure}[t]
\begin{center}
%\includegraphics[width=2.3in,height=2.6in]{fig_meynet3.eps}
%\hfill
\includegraphics[width=2.5in,height=2.2in]{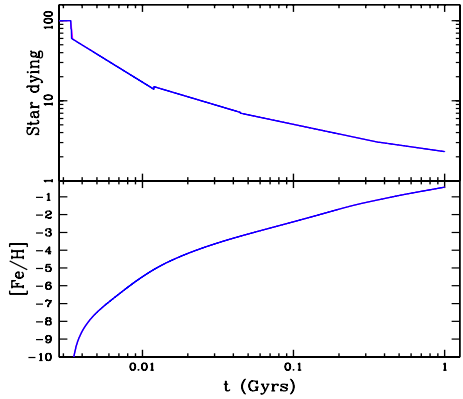}
\caption{
%\it Left panels}: 
After a starburst having occurred at time $t=0$, 
the top panel shows the
variation as a function of time of the minimum initial mass of
stars having finished their evolution at the considered time.
%; The
%evolution as a function of time of $Z$, the mass fraction of heavy elements,
%is plotted in the bottom panel.
%{\it Right panel}: 
%Same as left panel, 
The bottom panel shows the evolution as a function of time of [Fe/H].
The chemical evolution model is from Chiappini et al. (2006b).
}
\end{center}
\label{fig1}
\end{figure}

In the following, we shall call metallicity the quantity $Z$ which is the mass fraction
of all the elements with atomic mass number greater than that of helium.
In the course of the cosmic history, due to stellar nucleosynthesis, $Z$ increases globally in the Universe, although at different paces in different environments. In Fig.~\ref{fig1}, the evolution of
[Fe/H]
\footnote{
Note that [Fe/H] defined as
$\lg\left({(n_{Fe}/n_{H})_*\over (n_{Fe}/n_{H})_\odot}\right)$, where $n_{Fe}$, $n_{H}$
are the number density of iron, respectively of hydrogen and where the subscript $_*$ indicates
that the composition is that of the star considered and the subscript $_\odot$ that of the Sun.
} 
as predicted by the chemical evolution model for the halo of our Galaxy of Chiappini et al. (2006b) is shown. This evolution corresponds to that expected after an intense starburst which occurred at time zero and was responsible for driving a galactic wind.
One can see that 
%after about 0.4 Gyrs, $Z$ reaches values of about 0.002 and that 
ultra-metal poor stars ({\it i.e.} stars with [Fe/H] below -4) have acquired their
metallicity after less than 30 Myrs. Their heavy element content comes from  matter processed and ejected by stars with masses above 10 M$_\odot$.

At this stage it is interesting to note three points
\begin{itemize}
\item We note that metallicity increases very rapidly, indicating that the period of time
during which Pop III stars dominate the nucleosynthesis is very short. 
\item Systems like the halo of our Galaxy gives us the chance to observe stars whose composition
reflect only two specific nucleosynthetic processes: the Big Bang nucleosynthesis, and massive stars nucleosynthesis. This is in contrast with
for instance solar metallicity stars in the Milky Way whose initial heavy elements abundances reflect
the effects of many stellar generations and of various nucleosynthetic processes encompassing
{\it e.g.} intermediate mass stars, novae and 
type Ia supernovae. In that case,
the link between the observed
(surface) composition and the stellar yields is much less straightforward than for very metal poor halo stars. 
\item Finally, let us mention that a low or even very low metallicity does not necessarily implies that
only massive stars have contributed to the heavy elements enrichments of the stars. A very low metallicity can result from (very) low star formation rates over long periods.
In that case the surface abundances of non-evolved star reflect the stellar enrichments having occurred under the action of stars of
a much broader mass range than only massive stars. A way to discriminate between the case of a low star formation rate over long period and the one of a strong star formation rate observed at a very early stage is to look at abundance ratios (as for instance [$\alpha$/Fe], see Matteucci 2001).
In the case of the halo of our Galaxy the observed [$\alpha$/Fe] ratios indicates a short timescale for the heavy elements enrichment. 
\end{itemize}

The considerations above indicate that the halo of our Galaxy offers a wonderful opportunity to study
the nucleosynthesis due to the first massive star generations. As we shall see below many puzzling observed facts are related to the halo stars both in the field and in the globular clusters
(see Sect.~2.4 and 4). 

\section{The effects of metallicity on radiatively driven stellar winds and on rotational mixing}

Let us recall that a change of $Z$, the mass fraction of heavy elements, also affects
the initial mass fraction of hydrogen, $X$,  and of helium, $Y$. When $Z$ increases, $X$ decreases and $Y$ increases.
The increase of the abundance in helium with respect to $Z$ is conveniently described by a
quantity,
$\Delta Y/\Delta Z$, defined as the increase of helium per unit increase in $Z$.
The abundance of helium at a given stage, $Y$, is given by 
$Y=Y_0+\Delta
Y/\Delta Z\times Z$, where $Y_0$ is the primordial He abundance. A typical value
for $\Delta Y/\Delta Z$ at solar metallicity is around 2 (see e.g. Casagrande et al. 2007). Recently much higher values of
$\Delta Y/\Delta Z$ (of the order of 70) has been deduced from the 
presence in globular clusters of multi-ZAMS sequences (Piotto et al. 2005). We shall come back
on that question in Sect.~4 below. Thus probably
$\Delta Y/\Delta Z$ varies
as a function of the metallicity and of other environmental factors. 

At low metallicities
($Z\leq 0.02$),
$Y$ remains practically constant, and the
stellar properties {\it as a function of metallicity} are thus expected to be
determined mainly by $Z$. At high metallicities ($Z\geq 0.02$), on
the other hand, $Y$ increases (or, alternatively, $X$ decreases) significantly
with $Z$. In those conditions, both $Z$ {\it and} $Y$ (and $X=1-Y-Z$)
determine the stellar properties as a function of metallicity.  

In non rotating, non mass losing models, the metallicity affects the evolution of stars mainly through its
impact on the radiative opacities, the equation of state and the nuclear
reaction rates.
These effects of metallicity on stellar models are discussed in details in Mowlavi et al. (1998). 

For stars with initial mass greater than about 30 M$_\odot$, mass loss
becomes an important ingredient already during the MS phase and the effects of metallicity
on the mass loss rates have to be taken into account.
Metallicity also affects the transport mechanisms induced by rotation (Maeder \& Meynet 2001).
Typically a given initial mass star, starting its evolution with a given initial velocity, will be more efficiently mixed by rotation at low than at high metallicity. Below we
explain why such a behavior is expected.

\subsection{Line driven winds at different $Z$}

The main trigger of stellar winds is radiation pressure.
As written by Eddington (1926) `` {\it \dots the radiation observed to be emitted must work
its way through the star, and if there were too much
obstruction it would blow up the star.}'' 
It was realized already in the 20s that radiation pressure may produce mass loss. However it is only when, in the late 60s, sensitive UV diagnostics of mass loss from O-star became available that the effects of mass loss on the evolution of stars were really considered. 

Radiation triggers mass loss through the line opacities in hot stars. 
It may also power strong mass loss through the continuum opacity when the star is near the Eddington limit. For cool stars, radiation pressure is exerted on the dust.

For hot stars, typical values for the terminal wind velocity, $\upsilon_\infty$ is of the order
of 3 times the escape velocity, {\it i.e.} about 2000-3000 km s$^{-1}$,  mass loss rates are between 10$^{-8}$-10$^{-4}$ M$_\odot$ per year. Luminous Blue Variable (LBV) stars show during outbursts mass loss rates as high as 10$^{-1}$-1 M$_\odot$ per year!

According to the  recent mass loss rates, \mdot\ behaves with luminosity $L$
like $\mdot \sim L^{1.7}$.
With the mass--luminosity relation for massive stars, $L \sim \; M^2$, this gives
$\mdot \sim  M^{3.4}$.
From this we may  estimate the typical timescale for mass loss:
$t_{\mdot} \sim \frac{M}{\mdot } \sim \frac{1}{M^{2.4}}$.
This is to be compared to the MS lifetime
$t_{\mathrm{MS}}$. For massive stars, it scales like 
$t_{\mathrm{MS}} \sim  M^{-0.6}$
which shows that with increasing mass the timescale for mass loss decreases
much faster than the MS lifetime. One can also estimate the behavior of the
 amount $\Delta M$ of mass lost  with stellar mass 
$\Delta M  \sim M^{2.8} \quad \rightarrow \quad \frac{\Delta M}{M}  \sim  M^{1.8}$.
Thus, not only the amount of mass lost grows with the stellar mass, but even
the relative amount of mass loss grows fast with stellar mass, which illustrates the importance of this effect.

In addition to the intensity of the stellar winds for different evolutionary phases, one needs to know 
how the winds vary with the metallicity. This is a key effect to understand the different massive star populations observed in regions of different metallicities. This has also an important impact on the yields expected from stellar models at various metallicities (see Sect.~6).

Current wisdom considers that very metal-poor stars lose no or very small amounts
of mass through radiatively driven stellar winds. This comes from the fact that when the metallicity
is low, the number of absorbing lines is small and thus the coupling between the radiative forces and the matter is weak. Wind models impose a scaling relation of the
type 
\begin{equation}
\dot M(Z)=\left({Z \over Z_\odot} \right)^\alpha\dot M(Z_\odot),
\label{zmdot}
\end{equation}
where $\dot M(Z)$ is the mass loss rate when the metallicity is equal to $Z$ and $\dot M(Z_\odot)$
is the mass loss rate for the solar metallicity, $Z$ being the mass fraction of heavy elements.
In the metallicity range from 1/30 to 3.0 times solar,
the value of $\alpha$ is between 0.5 and 0.8 according to stellar wind models (Kudritzki et al. 1987; Leitherer et al. 1992; Vink et al. 2001). 
Such a scaling law implies for instance that 
a non-rotating 60 M$_\odot$ with $Z=0.02$
ends its stellar life with a final mass of 14.6 M$_\odot$, the same model with a metallicity of
$Z=0.00001$ ends its lifetime with a mass of 59.57 M$_\odot$ (cf. models of Meynet \& Maeder 2005 and Meynet et al. 2006 with $\alpha=0.5$). 

Thus
one expects that the metal-poor 60 M$_\odot$ star will give birth to a black hole.
In that case nearly all (if not all) the stellar mass may disappear in the remnant preventing the star from enriching in substantial way the
interstellar medium in new synthesized elements. The 
metal-rich model on the other hand will probably leave a neutron star and contribute to the enrichment of the ISM through both
the winds and the supernova ejecta. When the effects of rotation are accounted for, things can be very different at low metallicity. Mass loss can be boosted due to rotational effects (see Sect. 2.3, 2.4 and 4).

\subsection{Rotational mixing at different $Z$}

Many of the observed characteristics of massive stars require in order to be explained some extra-mixing mechanism working in their radiative zones (see e.g. the review by Maeder \& Meynet 2000a).
Rotation appears as the most promising mechanism for explaining these observed characteristics since massive stars are fast rotators and since many instabilities are triggered by rotation (Talon 2008). 
These instabilities transport angular momentum and chemical species in stellar interiors.
Assuming that
the star rapidly settles into a state of shellular rotation (constant angular velocity 
at the surface of isobars), the transport equations due to meridional currents and shear instabilities
can be consistently obtained (Zahn 1992). Since the work by J.-P.~Zahn, various improvements have been brought to the
formulas giving the velocity of the meridional currents (Maeder \& Zahn 1998), those of the various diffusive coefficients 
describing the effects of shear turbulence (Maeder 1997; Talon \& Zahn 1997; Maeder 2003; Mathis et al. 2004), as well as the effects of rotation on the mass loss (Owocki et al. 1996; Maeder 1999; Maeder \& Meynet 2000b). 

Let us recall a few basic results obtained from rotating stellar models (without dynamo process):

1) Angular momentum is mainly transported by the meridional currents. In the outer part
of the radiative envelope these meridional currents transport angular momentum outwards.
During the Main-Sequence phase, the core contracts and the envelope expands. The meridional currents
impose some coupling between the two, slowing down the core and accelerating the outer layers.
In the outer layers, the velocity of these currents becomes smaller when the density gets higher, {\it i.e.},
for a given initial mass, {\it when the metallicity is lower}.

\begin{figure}[t]
\includegraphics[width=2.3in,height=2.3in]{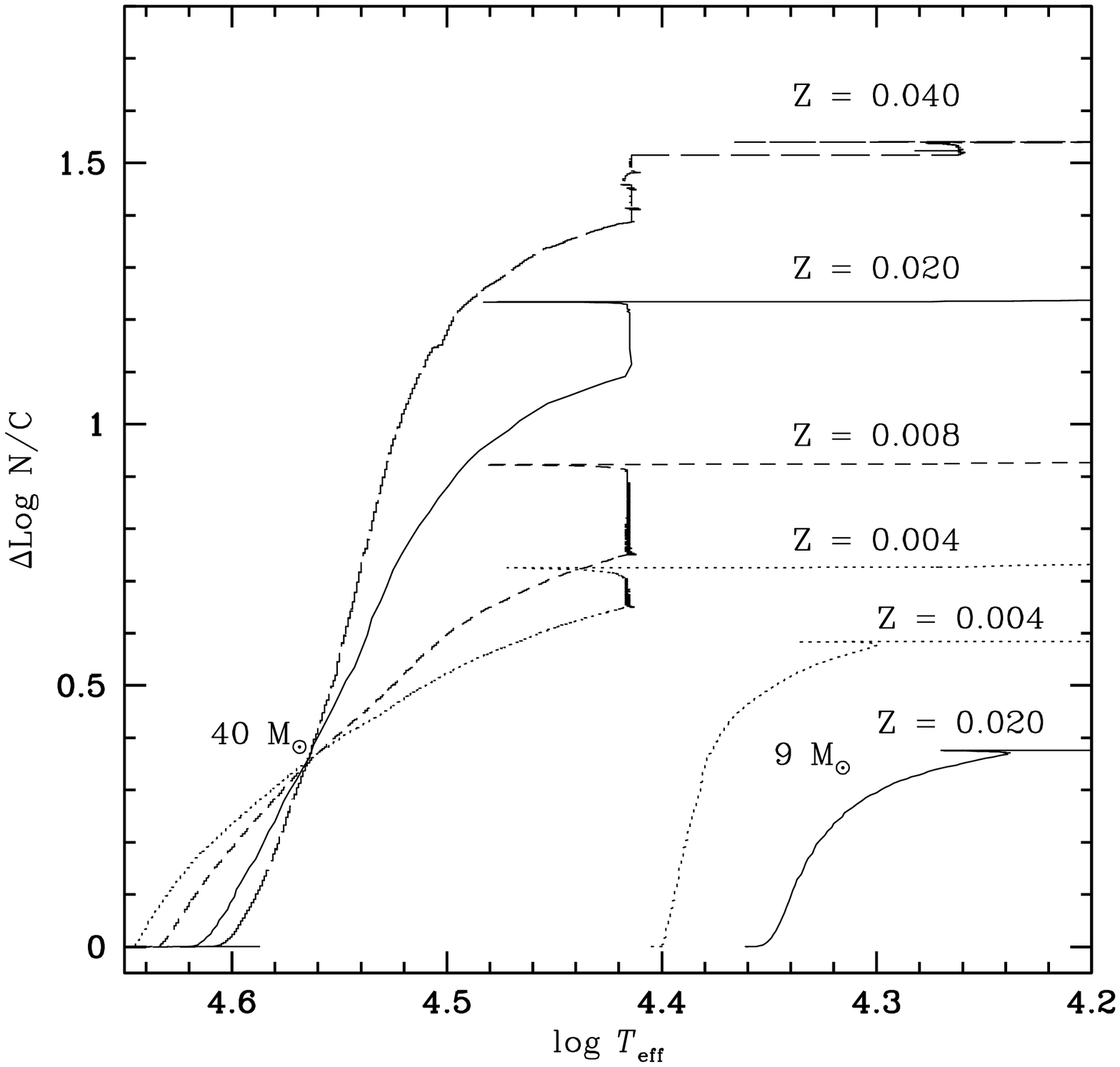}
\hfill
%\hspace{1cm}
\includegraphics[width=2.3in,height=2.3in]{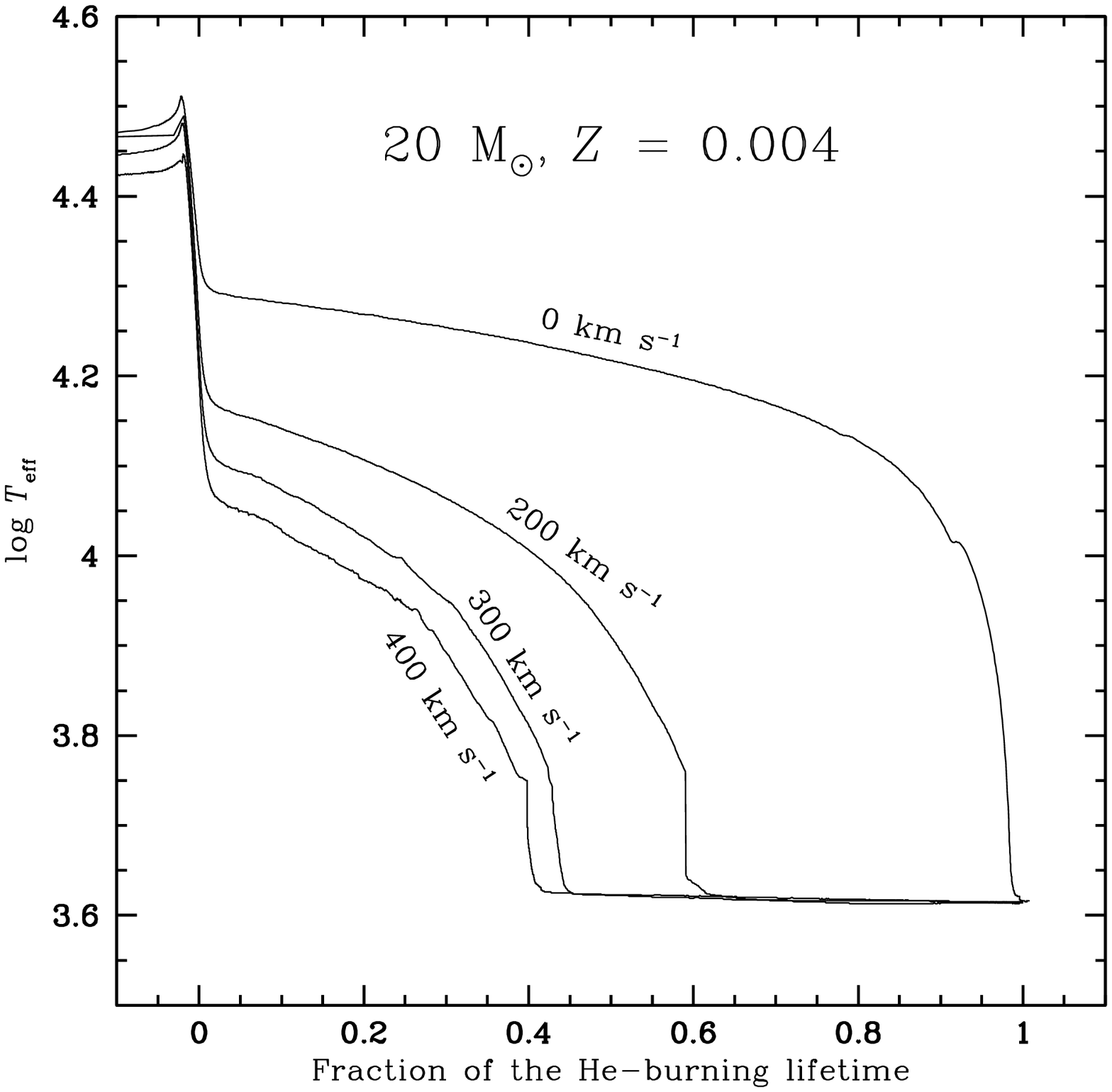}
\caption{{\it Left panel}: Evolution during the MS phase of the N/C ratios (in number) at the surface of rotating stellar models as a function of the effective temperature. 
The differences in N/C ratios are given with respect to the initial values.
Figure taken from  Meynet \& Maeder (2005).
{\it Right panel}: Evolution of the $T_{\mathrm{eff}}$
as a function of the fraction of the lifetime spent
in the He--burning phase for 20 M$_\odot$ stars with different
initial velocities. Figure taken from  Maeder \& Meynet (2001).}
\label{rSMC}
\end{figure}

2) The chemical species are mainly transported by shear turbulence\footnote{at least in absence of
magnetic fields; when magnetic fields are amplified by differential rotation as in the Tayler-Spruit
dynamo mechanism, see Spruit (2002), the main transport mechanism is meridional circulation, see Maeder \& Meynet (2005).}. 
During the Main-Sequence
this process is responsible for the nitrogen enhancements observed at the surface of most OB stars (Maeder et al. 2008). The shear turbulence is stronger when the gradients of the angular velocity are stronger. Due to point 1 above, the gradients of $\Omega$ are stronger in metal-poor stars and thus {\it the mixing of the chemical
elements is stronger in these stars}. This is illustrated on the left panel of Fig.~\ref{rSMC}
(see the tracks for the 9 M$_\odot$ stellar models).
Looking at the 40 M$_\odot$ stellar models, one sees that the higher metallicity model presents the highest surface enrichments, in striking contrast with the behavior of
the 9 M$_\odot$ model. This comes from the fact that
the changes occurring at the surface
of the 40 M$_\odot$ are not only due to rotation but also to mass loss which is more efficient at higher $Z$ (see also Sect.~6).
Mixing is not only more efficient at low metallicity, it is also stronger in stars with increasing initial masses and increasing initial velocities.

The efficiency of the mixing will vary from one element to another. If an element is strongly and rapidly built up in the convective core, it will diffuse by rotational mixing more rapidly in the radiative envelope than an element with a smoother gradient between the convective core and the radiative envelope. This explains
why the stellar surface will be more rapidly enriched in nitrogen than in helium. 

Some observations indicate that rotational mixing might be more efficient at lower metallicities (Venn 1999; Venn \& Przybilla 2003) confirming the trend expected from theoretical models.
However some variations in the initial distribution of the CNO elements might somewhat blurr the picture. According to Hunter et al. (2007) the C/N and O/N ratios in the Small Magellanic Clouds are equal to 7 and 35 respectively,
while in the solar neighborhood it is of the order of 4 and 7 (ratios in number). That means that
the maximum nitrogen enhancement (obtained at CNO equilibrium) expressed as $\Delta {\rm log(N/H)}$  would be about 1.7 in the SMC and about 1.1 in the solar neighborhood. Thus this effect
alone may produce higher nitrogen enrichment in the Magellanic Clouds than in the Galaxy!

In addition to these internal transport processes, rotation also modifies the physical properties
of the stellar surface. Indeed the shape of the star is deformed by rotation (a fact which is now put in evidence 
observationally thanks to the interferometry, see Domiciano de Souza et al. 2003). Rotation implies also a non-uniform brightness (see observational evidences in e.g. Domiciano de Souza et al. 2005).
The polar regions are brighter than the equatorial ones. This is a consequence of the hydrostatic
and radiative equilibrium (von Zeipel theorem 1924). In addition, as a result of the internal transport processes,
the surface velocity and the surface chemical composition are modified.

\subsection{Rotationally induced mass loss at different $Z$}

We can classify the effects of rotation on mass loss in three categories.

\begin{enumerate}
\item The structural effects of rotation.
\item The changes brought by rotation on the radiation driven stellar winds.
\item The mechanical mass loss induced by rotation at the critical limit ({\it i.e.} when
the surface velocity at the equator is such that the centrifugal acceleration balances the gravity).
\end{enumerate}

Let us consider in turn these various processes.
\vskip 2mm
\noindent\underline{Structural effects of rotation on mass loss}
\vskip 2mm
Rotation, by changing the chemical structure of the star, modifies
its evolution. For instance, moderate rotation at metallicities of the Small Magellanic Cloud (SMC)
favors redward evolution in the Hertzsprung-Russel diagram. This behavior is illustrated in the right panel of Fig.~\ref{rSMC} and can account for the high number of red supergiants observed in the SMC (Maeder \& Meynet 2001),
an observational fact which is not at all reproduced by non-rotating stellar models.

Now it is well known that the mass loss rates are greater
when the star evolves into the red part of the HR diagram, thus in this case, rotation modifies 
the mass loss indirectly, by changing the evolutionary tracks. 
The 20 M$_\odot$, $\upsilon_{\rm ini}=0$, 200, 300 and 400 km s$^{-1}$ models lose respectively 0.14, 1.40, 1.71 and 1.93 M$_\odot$ during the core He-burning phase (see Table~1 in Maeder \& Meynet 2001). The enhancement of the mass lost reflects the longer lifetimes of the red supergiant phase when velocity increases.
Note that these numbers were obtained assuming that the scaling law between mass loss and metallicity
deduced from stellar wind models for hot stars
applies during the red supergiant phase. If, during this phase, mass loss comes from continuum-opacity driven
wind then the mass-loss rate will not depend on metallicity (see the review by van Loon 2006).
In that case, the redward evolution favored by rotation would have a greater impact on mass loss than
that shown by these computations.

At very high rotation, the star will have a homogeneous evolution and will never become a red supergiant
(Maeder 1987).
In this case, the mass loss will be reduced, although this effect will be somewhat  compensated by two
other processes: first by the fact that the Main-Sequence lifetime will last longer and, second,
by the fact that the star will enter the Wolf-Rayet phase (a phase with high mass loss rates) at an earlier stage of its evolution. 

\vskip 2mm
\noindent\underline{Radiation driven stellar winds with rotation}
\vskip 2mm

\begin{figure}
\includegraphics[scale=0.3,angle=0]{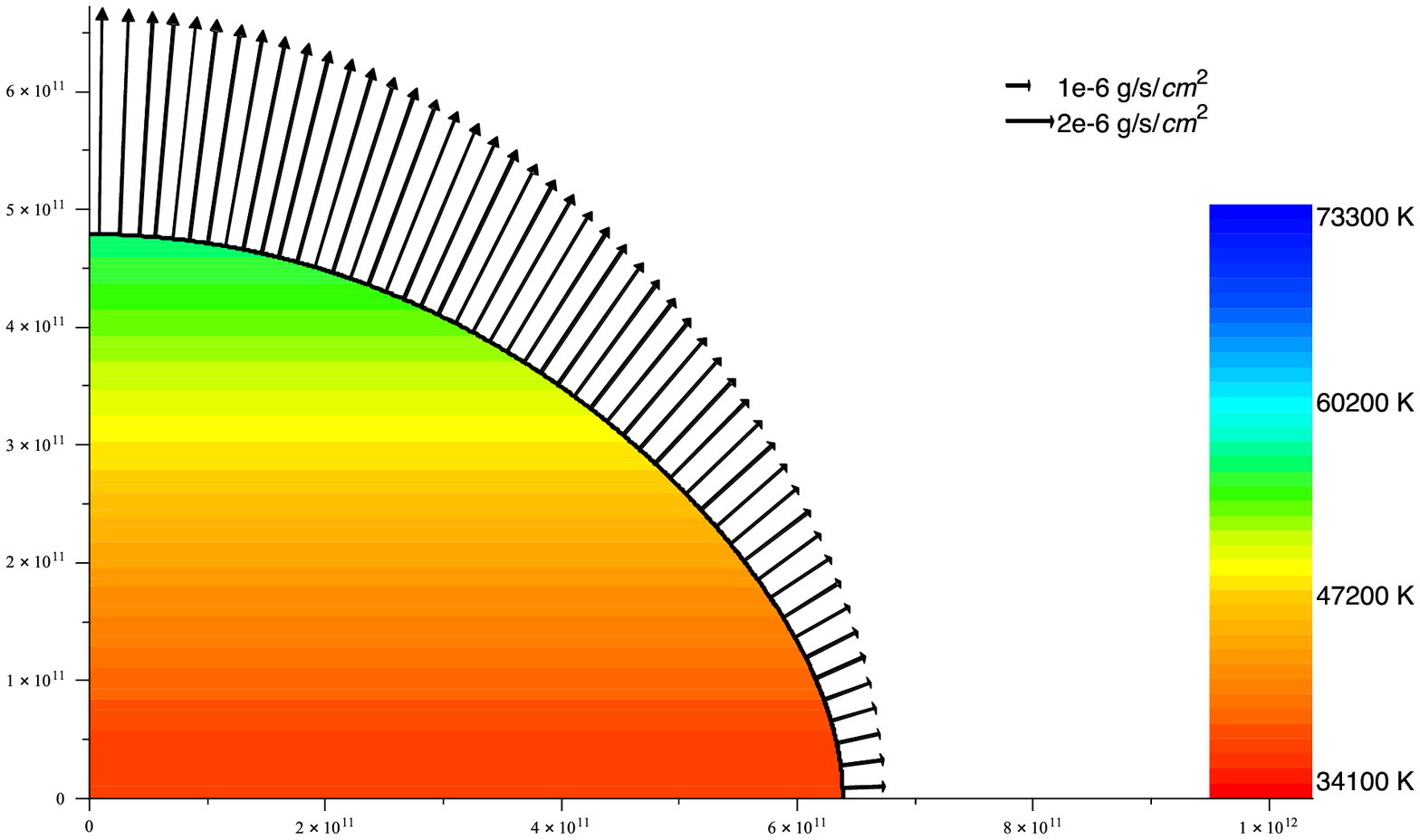}
\hfill
%\hspace{1cm}
\includegraphics[scale=0.3,angle=0]{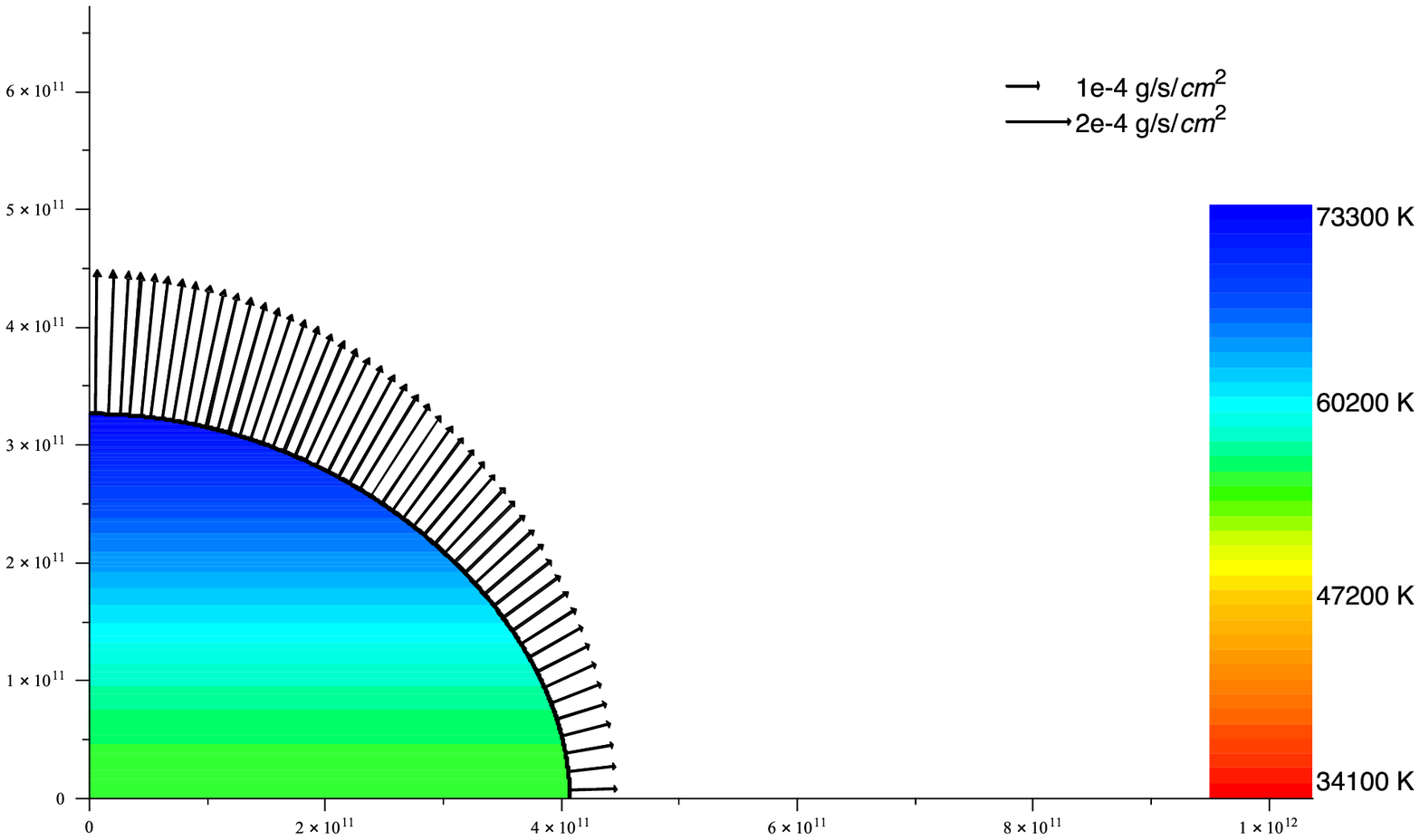}
\caption{{\it Left panel}: Variation of the mass flux at the surface of an initial 35 M$_\odot$ 
at a stage during the core H-burning phase when the mass fraction of hydrogen at the centre is 
$X_c=0.42$.
The velocity of the star on the ZAMS is 550 km s$^{-1}$ corresponding to $\Omega/\Omega_{\rm crit}=0.84$. The star follows a homogeneous evolution. At the stage represented  
$\Omega/\Omega_{\rm crit}\sim 1$. 
{\it Right panel}: Same as the left panel, but for a later stage with $X_c=0.02$ and  
$\Omega/\Omega_{\rm crit}\sim 1$.
%Same as the left panel, but for a later stage with $X_c=0.42$ and  
%$\Omega/\Omega_{\rm crit}~1$.
}\label{aniso1}
\end{figure}

The effects of rotation on the radiation driven stellar winds 
result from the changes brought by rotation to the stellar surface. They induce changes of the morphologies
of the stellar winds and increase their intensities.
\vskip 2mm
\noindent{\it Stellar wind anisotropies}
\vskip 2mm
Naively we would first guess that a rotating
star would lose mass preferentially from the equator, where the effective gravity (gravity decreased
by the effect of the centrifugal force) is lower.
This is probably true when the star reaches the critical limit (i.e. when the equatorial surface
velocity is such that the centrifugal acceleration exactly compensates the gravity), but this is not
correct when the star is not at the critical limit. Indeed as recalled above, a rotating star has a
non uniform surface brightness, and the polar regions are those which have the most powerful radiative 
flux. Thus one expects that the star will lose mass preferentially along the rotational axis. This is
correct for hot stars, for which the dominant source of opacity is electron scattering. In that
case the opacity only depends on the mass fraction of hydrogen and does not depend on other physical quantities such as temperature. Thus rotation induces 
anisotropies of the winds  (Maeder \& Desjacques 2001; Dwarkadas \& Owocki 2002).
Amplitude of the effect for a fast rotating 35 M$_\odot$ stellar model can be seen in Fig.~\ref{aniso1}.

Wind anisotropies have consequences for the angular momentum that a star retains in its interior.
Indeed, when mass is lost preferentially along the polar axis, less angular momentum is lost.
This process allows loss of mass without too much loss of angular momentum, a process which might
be important in the context of the evolutionary scenarios leading to Gamma Ray Bursts (GRB). Indeed 
in the framework of the collapsar scenario (Woosley 1993), 
one has to accommodate two contradictory requirements: on one side, the progenitor needs to lose mass
in order to have its H and He-rich envelope removed at the time of its explosion\footnote{In case the H-rich envelope would be still present at the time of the explosion, the jets and the gamma photons produced in it would remain invisible for an observer.}, and on the other hand it must have retained sufficient angular momentum in its central region to give birth to a fast
rotating black-hole. Wind anisotropies may help a fast rotating star to lose mass without losing too much angular momentum.

\vskip 2mm
\noindent{\it Intensities of the stellar winds}      
\vskip 2mm      
The quantity of mass lost through radiatively driven stellar winds is enhanced by rotation. This enhancement can occur through two channels: by reducing the effective gravity at the surface of the star, by increasing the opacity of the outer layers through surface metallicity enhancements due to rotational mixing.
      
\begin{itemize}

\item{\it reduction of the effective gravity: } The ratio of the mass loss rate of a star with a surface angular velocity $\Omega$ to that
of a non-rotating star, of the same initial mass, metallicity and lying at the same position in the
HR diagram is given by (Maeder \& Meynet 2000b)
      
\begin{equation}
\frac{\dot{M} (\Omega)} {\dot{M} (0)} \simeq
\frac{\left( 1  -\Gamma\right)
^{\frac{1}{\alpha} - 1}}
{\left[ 1 - 
\frac{4}{9} (\frac{v}{v_{\mathrm{crit, 1}}})^2-\Gamma \right]
^{\frac{1}{\alpha} - 1}} \; ,
\end{equation}
\noindent
where $\Gamma$ is the electron scattering opacity for a non--rotating
star with the same mass and luminosity, $\alpha$ is a force multiplier (Lamers et al. 1995). 
The enhancement factor remains modest for stars with luminosity sufficiently far away from the
Eddington limit (Maeder \& Meynet 2000b). Typically, $\frac{\dot{M} (\Omega)} {\dot{M} (0)} \simeq 1.5$ for main-sequence B--stars.
In that case, when the surface velocity approaches the critical limit, the effective
gravity decreases and the radiative flux also decreases. Thus the matter becomes less bound
when, at the same time, the radiative forces become also weaker. 
When the stellar luminosity approaches the Eddington limit, the mass loss increases can be much greater, reaching orders of magnitude.
This comes from the fact that rotation lowers the maximum luminosity or the Eddington luminosity of a star.  Thus it may happen that for a velocity still 
far from the classical critical limit, the 
rotationally decreased maximum luminosity becomes equal to the actual luminosity of the star. 
In that case, strong mass loss ensues and the star is said to have reached
the $\Omega\Gamma$ limit (Maeder \& Meynet 2000b).

\item {\it Effects due to rotational mixing: }
During the core helium burning phase, at low metallicity,
the surface may be strongly enriched in both H-burning and He-burning products, {\it i.e.} mainly in nitrogen, carbon and oxygen. Nitrogen is produced by transformation of the carbon and oxygen produced in the He-burning core and which have diffused by rotational mixing in the H-burning shell (Meynet \& Maeder 2002ab). Part of the carbon and oxygen produced in the He-core also diffuses up to the surface. Thus at the surface, one obtains very high value of the CNO elements, the opacity of the surface increases and thus line driven winds become stronger (see Sect.~4 for quantitative
effects).
%For instance a 60 M$_\odot$ with Z=$10^{-8}$ and $\upsilon_{\rm ini}=800$ km s$^{-1}$ has, at the end of its evolution, a CNO content at the surface equivalent to 1 million times its initial metallicity! In the present models, we have applied the usual scaling laws linking the surface metallicity
%to the mass loss rates (see Eq.~\ref{zmdot}). In that case, one obtains that the star loses 
%more than half of its initial mass due to this process.

\end{itemize}

\vskip 2mm      
\noindent\underline{Mechanical winds induced by rotation}      
\vskip 2mm
      
As recalled above, during the Main-Sequence phase the core contracts
and the envelope expands. In case of local conservation of the angular momentum, the core would thus
spin faster and faster while the envelope would slow down. In that case, it can be easily shown that the surface velocity would evolve away from the critical velocity (see e.g. Meynet \& Maeder 2006). 
In models with shellular rotation however
an important coupling between the core and the envelope is established through the action of the
meridional currents. As a net result, angular momentum is brought from the inner regions to the outer ones. Thus, would the star lose no mass by radiation driven stellar winds (as is the case at low Z), one expects that the surface velocity
would increase with time and would approach the critical limit. In contrast, 
when radiation driven stellar winds are important, the timescale for removing mass 
and angular momentum at the surface
is shorter than the timescale for accelerating the outer layers by the above process and the surface velocity decreases as a function of time. It evolves away from the critical limit. 
Thus, an interesting situation occurs: when the star loses
little mass by radiation driven stellar winds, it has more chance to lose mass by a mechanical wind. On the other hand, when the star loses mass at a high rate by
radiation driven mass loss then it has no chance to reach the critical limit and thus to undergo a 
mechanical mass loss. 

\subsection{Spinstars at low metallicity?}

From the previous discussion, it does appear that for a given initial rotation, the effects of rotation will be the more pronounced at low metallicity and this for two reasons:
\begin{itemize}
\item Rotational mixing is more efficient at low $Z$.
\item Stars more easily evolves toward the critical limit.
\end{itemize}
Of course if at low metallicity, for whatever reasons, the distribution of the initial velocities
is biased toward slow rotators, then the above effects will not be important. Let us however
stress here that at present, there is some observational support to the view that the distribution of the initial velocities contains more fast rotators at low Z! Observationally Martayan et al (2006) find that, for B and Be stars, the lower the metallicity, the higher the rotational velocities.
The Be stars are stars surrounded by an expanding equatorial disks probably
produced by the concomitant effects of both fast rotation and pulsation.
Martayan et al. (2006; 2007) obtain that Be stars rotate faster than B stars whatever the metallicity is. Thus a high fraction of Be stars reflect a higher fraction of fast rotators. 
Maeder et al. (1999) and Wisniewski \& Bjorkman (2006)
find that the fraction of Be stars with respect to the total number of B and Be stars
in clusters with log t(yrs) from 7.0 to 7.4 
increases when the metallicity decreases. This fraction passes from about 10\% at solar metallicity to about 35\% at the SMC metallicity. 

Thus not only for a given initial velocity, theory predicts that the effects of rotation are more pronounced in metal poor regions but also there is some observational hints that there are
more fast rotators at low $Z$. Of course we might argue here that the higher proportion of fast rotators results not from a difference in the initial distribution of the velocities but is due to
an evolutionary effect. Stars at low $Z$ rotate faster because they lose less angular momentum through radiatively driven stellar winds. However for B-type stars, stellar winds are weak during the
Main-Sequence phase and this argument probably does not apply.
%\footnote{We do not consider here the possibility that magnetic braking is important in that case. Would this mechanism operate then even
%a very low mass loss may bring away large amount of angular momentum.}.

From this we conclude that rotation probably has the strongest impact at low metallicity. If
we define ``spinstars'' as stars whose evolution is strongly affected by rotation, then their
number should be greater at low $Z$. 

Interestingly many observations presently challenge our understanding of the evolution of stars at low and very low metallicity. Let us briefly
mention a few of them here: 1) no sign of Pair Instability Supernovae has been observed (Cayrel et al. 2004), 2) a high plateau of the  N/O ratios (as a function of O/H) are obtained requiring the activity  of efficient sources of primary nitrogen (see e.g. Spite et al. 2005), 3) simultaneously the C/O ratio as a function of O/H shows an upturn at a metallicity of [O/H]$\sim$-2  (Cayrel et al. 2004), similar trends are seen in Damped Lyman Alpha systems (Pettini et al. 2008) 4) a significant fraction of very metal poor stars are C-rich stars showing very peculiar abundance pattern at their surface (see the review by Beers \& Christlieb 2005), 5) very helium-rich stars and 6) stars with high abundance of sodium and low abundance of oxygen  are detected in globular clusters (Gratton et al. 2004; Piotto et al. 2005).
As we shall see below rotation may help in understanding these puzzling facts.

\section{The first stellar generations in the Universe}

Understanding the evolution of massive stars at low and very low metallicity is a requirement to address questions such as the nature of the sources of the reionization in the early Universe, 
the evolution of the interstellar abundances during the early phases of the evolution of galaxies, for finding possible signatures of primordial stellar populations in the integrated light of
very distant galaxies and for discovering which objects are the progenitors of the long soft Gamma Ray Bursts. At present, the most ``iron'' poor objects known in the Universe are not very far from us since they are galactic halo field stars. These objects offer a unique opportunity to study the yields of the first generations of stars\footnote{Note that recently Venn \& Lambert (2008) have challenged the view according to which the C-rich extremely metal poor stars are
trustworthy very metal poor stars. But this suggestion, if correct, will not challenge the whole concept of the  first stellar generations being traced by metal poor stars because many of the chemical signatures also come from normal stars ({\it i.e.} non C-rich stars).}.

In this section, we focus the discussion on 
the first stars which were made up of material having been processed only by 
primordial nucleosynthesis, that means from matter essentially deprived of heavy elements.
%\footnote{Standard Big Bang nucleosynthesis predicts that the amount of CN elements
%produced is of the order of 10$^{-13}$ in mass fraction (value obtained by Volanthen 2003 using
%the code of Thielemann).}. 
In an environment with primordial composition, one expects the following differences with respect
to the more classical evolution at higher metallicities:
\begin{itemize}
\item At strictly $Z=0$, the cooling processes, so important for allowing the
evacuation of the energy produced when the molecular clouds collapse and thus
its fragmentation, are not so efficient as at larger metallicity. This favors the formation of more
massive stars. The initial mass function is probably different depending on the mass
of the ``minihaloes'' (see Greif \& Bromm 2006). In minihaloes with masses between 10$^6$ and
10$^8$ M$_\odot$, virial temperature is between 10$^3$ and 10$^4$ K and the cooling is due mainly to
the molecules H$_2$. This allows the formation of stars with characteristic masses $\ge 100$ M$_\odot$. In minihaloes where ionization occurs prior to the late stages of the protostellar accretion process, namely those with a virial temperature superior to 10$^4$ K and thus with masses
above 10$^8$ M$_\odot$\footnote{This may also happen in relic HII regions left by the first stars, see Johnson \& Bromm (2006).}, the hydrogen deuteride (HD) molecule provides an additional cooling channel.
In those minihaloes, metal-free gas can cool more efficiently. This leads to the formation of stars
with masses superior to 10 M$_\odot$. Thus at the very beginning, we would have first, during a quite
short period, only very massive Pop III stars and, when gas assembles in more massive haloes (or is
reionized by the first stars), a second Pop III star generation appears with smaller characteristic masses (about 10 M$_\odot$ and above). This population is called by some authors Pop II.5.
When the
metallicity becomes higher than 10$^{-3.5}$ Z$_\odot$, {\it i.e.} for $Z$ above about 5 10$^{-6}$
a normal IMF governs the mass distribution of newly born stars\footnote{Some authors argue that
star formation switches to more classical mode already when $Z=10^{-6}$ due to dust production in the early Universe, see Schneider et al. 2006.}. According to Greif \& Bromm (2006) the very massive Pop III stars only contribute marginally to feed the reservoir of ionizing photons and to the chemical enrichment of the interstellar medium. Much more important are on this respect the less massive Pop III stars born in more massive haloes or in relic HII regions.
\item The (nearly) absence of heavy elements implies that massive stars cannot ignite hydrogen-burning through the CNO cycle but through the pp chains. However the energy output extracted from the pp chains is not sufficient to compensate for the high luminosity of these stars. The stars must compensate the deficit  of nuclear energy by extracting energy from the gravitational reservoir, {\it i.e.} they contract. At a given point however, due to this contraction, the central temperature becomes high enough for activating triple $\alpha$ reactions. Some carbon is then produced. When the mass fraction of carbon is of the order of 10$^{-12}$, the CN cycle becomes the main source of energy and the evolution then proceeds as in a more metal rich massive stars. For stars above about 20 M$_\odot$, activation of the CN cycle intervenes very early during the core H-burning (typically before five percents of the hydrogen at the centre is consumed, see Marigo et al. 2003).
\item Pop III are more compact due to the contraction they undergo at the beginning of the core H-burning phase (see above) and to the fact that the opacity of primordial material is smaller than that of metal rich one. Typically a Pop III 20 M$_\odot$ star on the ZAMS has a radius reduced by a factor 3.5 with respect to the radius of the corresponding star with $Z=0.020$. Even passing
from the very low metallicity $Z=10^{-5}$ to 0 already produces a decrease of the radius by a factor 2! Smaller radius favors more efficient mixing in two ways: first, for a given value of the diffusion coefficient, $D$, the mixing timescale decreases with the radius as $\sim R^2/D$. Second the gradients of $\Omega$ are steeper (see above).
As we shall see however, rotational mixing at $Z=0$ is not so efficient as at low but non zero metallicity. We shall come back on that point in the next section.
\end{itemize}

\begin{figure}[t]
\includegraphics[width=2.3in,height=2.3in]{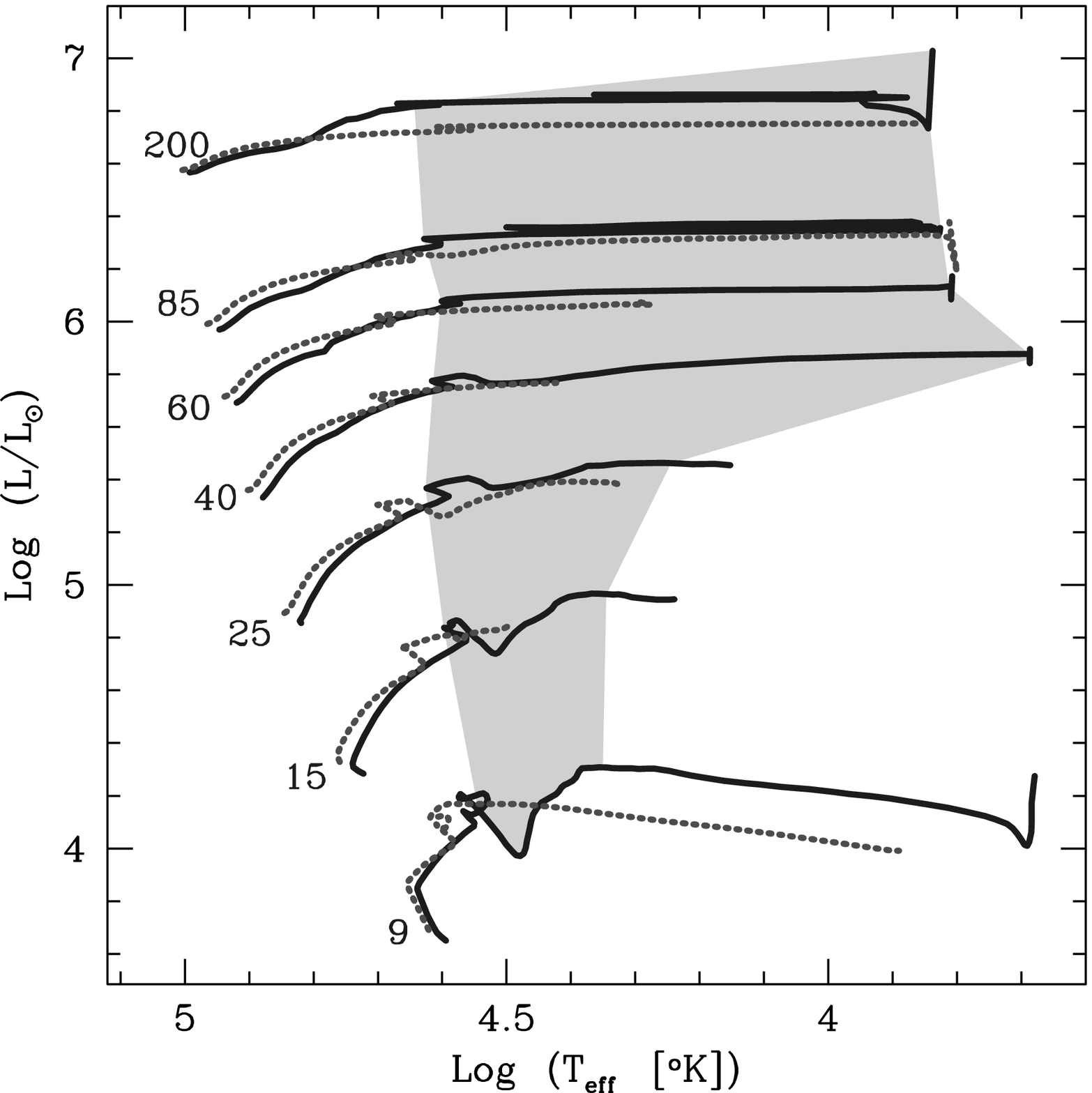}
\hfill
\includegraphics[width=2.3in,height=2.3in]{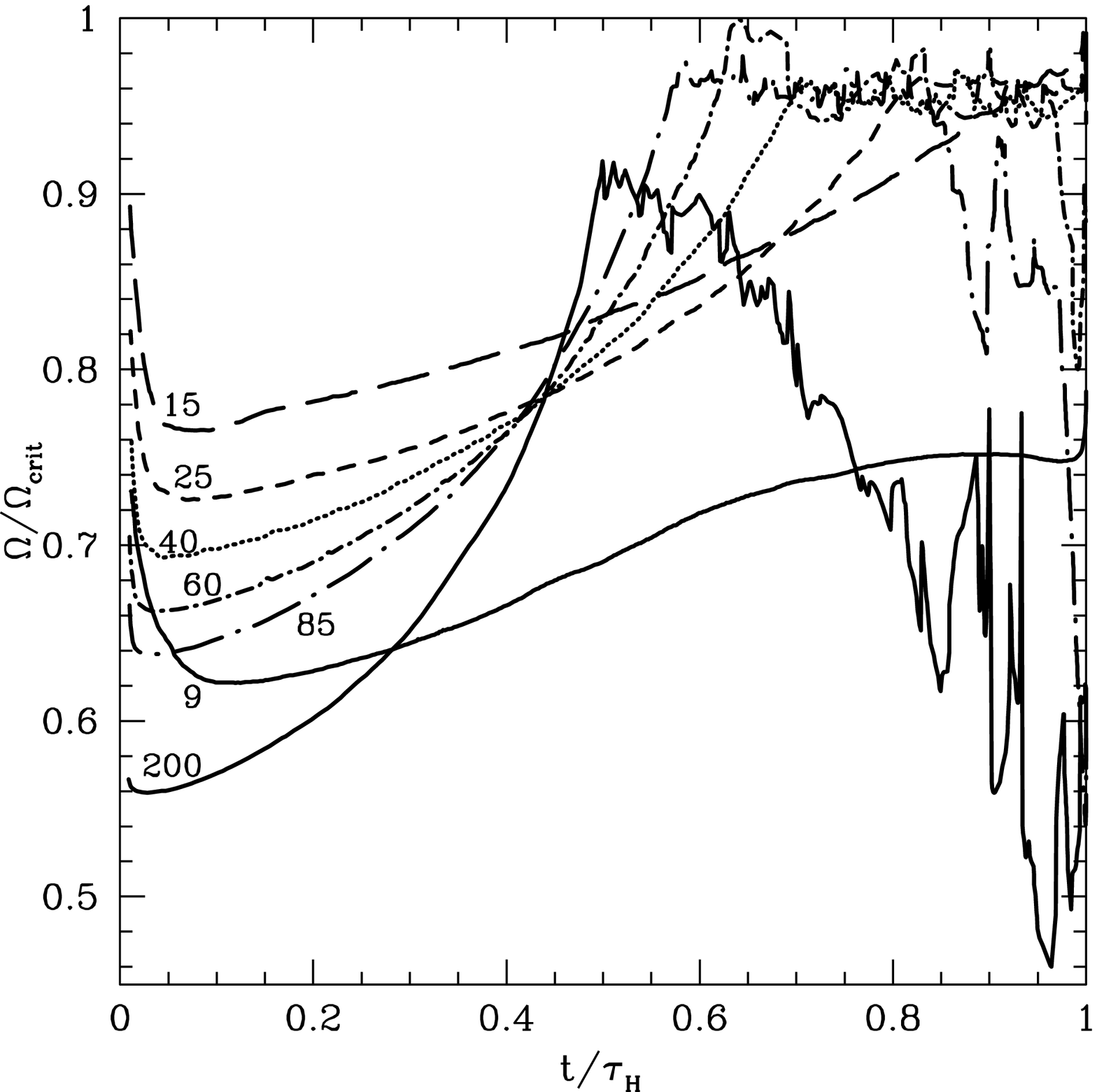}
\caption{
{\it Left panels}: Evolution of $Z=0$ models (with rotation: continuous lines; without rotation: dotted lines) in the Hertzsprung Russell diagram. The gray area shows the zone of the diagram where He burns in the core of the rotating models.
{\it Right panel}: Evolution of the $\Omega/\Omega_\mathrm{crit}$ ratio during the MS. All the models start the MS with $\upsilon_\mathrm{eq}=800$ km s$^{-1}$, except the 9  M$_\odot$ which starts the MS with 500 km s$^{-1}$. Figures taken from Ekstr\"om et al. (2008a).
}
\label{fig2}
\end{figure}

Recently Ekstr\"om et al. (2008) presented a grid of Pop III stellar models including the effects of rotation. 
Evolutionary tracks of non-rotating and rotating Pop III stellar models are shown in Fig.~\ref{fig2}
(left panel). The models are computed until the end of the core Si-burning, except the 9 M$_\odot$ that has developed a degenerate core before carbon ignition and has thus been stopped then, and the 15 M$_\odot$ model that has been stopped at the end of O-burning also because of a too degenerate core at that time. We chose an initial velocity of 800 km s$^{-1}$ on the ZAMS.
For the 60 M$_\odot$ model, this corresponds to a value of $\upsilon/\upsilon_{\rm crit}=0.52$ on the ZAMS, which is slightly superior to 0.4, the value required at solar metallicity to obtain averaged velocities during the MS phase corresponding to observed values. 

We notice that the ZAMS is shifted toward lower effective temperature and luminosity with respect to the non-rotating case\footnote{We recall that this shift is due to the sustaining effect of rotation: the gravity is counter-balanced both by the gas pressure and the centrifugal force in such a way that the star behaves like a lower mass one.}. Then, when the evolution proceeds, the tracks become more luminous, and the main-sequence turn-off is shifted to cooler temperature: the core of the rotating models is refueled by fresh H brought by the mixing. It thus grows, leading to an enhancement of the luminosity.

The onset of the CNO cycle described above can be seen in the HRD:  the tracks evolve toward the blue side of the diagram, until the energy provided by the CNO cycle stops the contraction and bends the tracks back in the usual MS feature. In the rotating 9 M$_\odot$, this happens at an age of 12.2 Myr (when the central H mass fraction is $X_\mathrm{c}=0.439$) while in the non-rotating one it happens a little earlier, at an age of 10.9 Myr (but at a similar burning stage: $X_\mathrm{c}=0.439$). In the case of the non-rotating 15 M$_\odot$ model, it happens after merely 1.5 Myr ($X_\mathrm{c}=0.695$), while it takes 2 Myr ($X_\mathrm{c}=0.677$) in the case of the rotating one. Let us mention that Marigo et al. (2003) find that the mass limit for CNO ignition already on the ZAMS is 20 M$_\odot$, our results being consistent with that limit.

After central H exhaustion, the core He-burning phase (CHeB) starts right away: the core was already hot enough to burn a little He during the MS and does not need to contract much further.
This prevents the models to start a redward evolution, so they remain in the blue part of the HRD at the beginning of CHeB. Then, something particular happens to the rotating models: because of rotational mixing, some carbon produced in the core is diffused toward the H-burning shell, allowing a sudden ignition of the CNO-cycle in the shell. This boost of the shell leads to a retraction of the convective core and a decrease of the luminosity. At the same time, it transforms the quiet radiative H-burning shell into an active convective one. Some primary nitrogen is produced
(see more on that point in the next section). 

All the models, except the 9 M$_\odot$, reach the critical velocity during the MS phase (see the right panel of Fig.~\ref{fig2}). Once at critical limit, all the models remain at the critical limit until the end of the Main-Sequence phase. On the right panel of Fig.~\ref{fig2}, the 85 and especially the 200 M$_\odot$ seem to depart from $\Omega/\Omega_\mathrm{crit}=1$, but this is due to the limit shown here being only the $\Omega$-limit, where the centrifugal force alone is taken into account to counterbalance the gravity. In the two above models however, the radiative acceleration is strong and the models reach the so-called $\Omega\Gamma$-limit, that is the second root of the equation giving the critical velocity: $\vec{g_\mathrm{eff}}\,[1-\Gamma]=\vec{0}$ (Maeder \& Meynet 2000b). The true critical velocity is lowered by the radiative acceleration, and though the $\Omega/\Omega_\mathrm{crit}$ ratio plotted becomes lower than 1, these models are actually at the critical limit and remain at this limit till the end of the MS. 
The mass which is lost by the mechanical winds amounts only to a few percents of the initial stellar mass and thus does not
much affect neither their evolution, nor their nucleosynthetic outputs. Much more mass can be lost by mechanical mass loss (see next subsection) when the effects of magnetic fields are accounted for as prescribed in the Tayler-Spruit dynamo theory (Spruit 2002), or when the metallicity is non-zero.

\subsection{Strong mass loss in Pop III stars?}

\begin{figure}[t]
% \vspace*{-2.0 cm}
\begin{center}
 \includegraphics[width=0.8\textwidth]{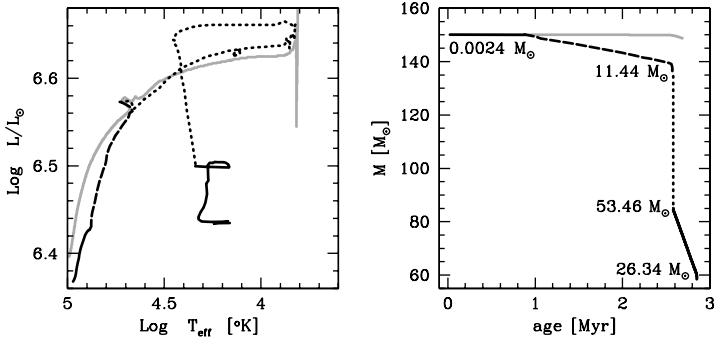}
% \vspace*{-1.0 cm}
 \caption{Black line: rotating model; \emph{continuous part}: beginning of MS ($X_{\rm c}=0.753$ down to 0.58; \emph{dashed part}: rest of the MS; \emph{dotted part}: beginning of core He-burning phase ($Y_{\rm c}=1.00$ down to 0.96); \emph{continuous part}: rest of the He-burning. Grey line: non-rotating model for comparison. \textbf{Left panel}: evolution in the Hertzsprung-Russell diagram; \textbf{Right panel}: evolution of the mass of the model. The mass indicated is the mass lost at each stage, not a summation. Figures taken from Ekstr\"om et al. (2008b).}
   \label{fevol150}
\end{center}
\end{figure}

According to Heger et al. (2003), the fate of single stars depends on their He-core mass ($M_{\alpha}$) at the end of the evolution. They have shown that at very low metallicity, the stars having $64\ M_{\odot} < M_{\alpha} < 133\ M_{\odot}$ will undergo pair-instability and be entirely disrupted by the subsequent supernova. This mass range in $M_{\alpha}$ has been related to the initial mass the star must have on the main sequence (MS) through standard evolution models: $140\ M_{\odot} < M_{\rm ini} < 260\ M_{\odot}$. However the link between the masses of the He-core mass and the initial mass
can be very different depending on the physics considered. Ekstr\"om et al. (2008) showed that
a Z=0, 150 M$_{\odot}$ stellar model, having an initial ratio between the equatorial velocity and the critical one equal to $\upsilon_{\rm ini}/\upsilon_{\rm crit} = 0.56$, and computed accounting
for the Tayler-Spruit dynamo mechanism (Spruit 2002) and the effects of wind anisotropy (Maeder 1999) will lose such great amount of mass that it will reach the end of the core He-burning phase with a mass of $M_{\alpha}$ too small to go through a pair-instability process.

In Fig. \ref{fevol150}, we present the evolution in the HR diagram (left panel) and the evolution of mass with time (right panel). The gray line shows a non-rotating model computed with the same physics for comparison.
During its whole evolution up to the end of core He-burning, the non-rotating model loses only $1.37\ M_{\odot}$. This illustrates the weakness of radiative winds at $Z=0$.
The evolution of the rotating model (black line) can be described by four distinct stages:
\begin{enumerate}
\item \emph{(continuous part}, lower left corner in the left panel of Fig.~\ref{fevol150}) The model starts its evolution on the MS with only radiative winds, losing only a little more than $0.002\ M_{\odot}$. During this stage, the ratio
of the surface velocity to the critical one increases quickly, mainly because of the strong coupling exerted by the magnetic fields.

\item \emph{(dashed part)} When its central content of hydrogen is still 0.58 in mass fraction, it reaches the critical velocity and starts losing mass by mechanical mass loss. It remains at the critical limit through the whole MS, but the mechanical wind removes only the most superficial layers that have become unbound, and less than 10\% of the initial mass is lost during that stage ($11.44\ M_{\odot}$). The model becomes also extremely luminous, and reaches the Eddington limit when 10\% of hydrogen remains in the core. Precisely, it is the so-called $\Omega\Gamma$-limit that is reached here. 

\item \emph{(dotted part)} The combustion of helium begins as soon as the hydrogen is exhausted in the core, then the radiative H-burning shell undergoes a CNO flash, setting the model on its redward journey. The model remains at the $\Omega\Gamma$-limit and loses a huge amount of mass. The strong magnetic coupling keeps bringing angular momentum to the surface and even the heavy mass loss is not able to let the model evolve away from the critical limit. The mass lost during that stage amounts to $53.46\ M_{\odot}$. When the model starts a blue hook in the HR diagram, its surface conditions become those of a WR star ($X_{surf} < 0.4$ and $T_{\rm eff} > 10'000$ K). The luminosity drops and takes the model away from the $\Gamma$-limit, marking the end of that stage.

\item \emph{(continuous part)} The rest of the core He-burning is spent in the WR conditions. The mass loss is strong but less than in the previous stage: another $26.34\ M_{\odot}$ are lost.
\end{enumerate}

At the end of core He-burning, the final mass of the model is only $M_{\rm fin}=58\ M_{\odot}$, already below the minimum $M_{\alpha}$ needed for PISN ($M_{\alpha} \geq 64\ M_{\odot}$). Note that the contraction of the core after helium exhaustion brings the model back to critical velocity, so this value for $M_{\rm fin}$ must be considered as an upper limit.

This result shows that a fast rotating Pop III 150 M$_\odot$ may avoid to explode as a PISN. Also
such a star will enrich the interstellar medium through its winds.
Of course it is by far not certain that the conditions required for such a scenario to occur are met in the first stellar generations but it underlines the fact that fast rotation may drastically change the picture. 

Let us note that the nucleosynthetic signature of PISN are not observed in the most metal poor halo stars. Is this due to the above scenario? To the fact that the signature was very quickly erased by the next generations of stars?\footnote{Maybe the metal-poor stars we observe are enriched by more SNe than we actually think, and the later contributions are masking the primordial ones.} Or were such high mass stars not formed? These various hypothesis cannot be disentangled at the present time, but the observation of more and more metal-deficient stars will probably provide elements of response to these questions.

\subsection{Pop III stars as physics laboratories}

The formation and the evolution of the first stellar generations may differ from those of the subsequent generations in many other different ways than the ones quoted above. For instance Pop III stars are supposed to be formed in mini dark haloes. Depending on the nature of dark matter, of its density in the dark halo,
annihilation of dark material inside Pop III stars may contribute 
more or less significantly for compensating their energy losses at the surface.

\begin{figure}[t]
\begin{center}
\includegraphics[width=2.5in,height=2.5in]{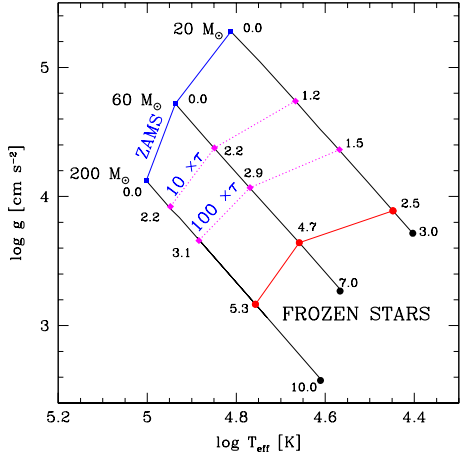}
%\hfill
%\includegraphics[width=2.3in,height=2.6in]{XicM015_endHe.png}
\caption{
ZAMS positions of 20, 60 and 200 M$_\odot$ Pop III stellar models
in the surface gravity versus effective temperature plane (both in logarithm) for
different dark matter densities (labels in units of GeV cm$^{-3}$). Big red (gray) circles
labeled by 5.3, 4.7 and 2.5, correspond to dark matter densities such that 
the energy produced by WIMPS annihilation can compensate for the energy losses at the surface.
Provided such densities can be sustained for more tan a Hubble time, the star
do no long evolve and remains frozen at its ZAMS position. The lines labeled as 10$\tau$
and 100$\tau$ correspond to ZAMS positions of models with a lifetime prolonged by 10 and 100
times with respect to the case without WIMPS. We have adopted a WIMP model with a mass equal
to 100 GeV and a spin-dependent WIMP scattering cross section  $\sigma_{\rm SD}=10^{-38}$ cm$^{2}$.
Figure taken from Taoso et al. (2008).
%{\it Right panel}: Chemical composition of the He-burning core at the end of the core He-burning phase for different values of the binding energy of deuterium (from Ekstr\"om et al. in preparation).
}
\end{center}
\label{dmtao}
\end{figure}

Recently many authors (see e.g. 
Spolyar et al. 2008; Iocco et al. 2008; Taoso et al. 2008; Yoon et al. 2008)  have studied the effects of dark matter annihilation in Pop III stars. In particular Taoso et al. (2008) have obtained that for WIMP densities superior to a critical value (see Fig.~\ref{dmtao}) the MS lifetimes of these models can exceed the present age of the Universe, allowing the existence in the present day Universe of Pop III stars still frozen in their ZAMS evolutionary stage. In contrast to 
classical ZAMS Pop III stars, these stars would be much more inflated, show lower effective temperatures and surface gravities. The circumstances needed for such stars to exists are probably very rare if ever they have been realised once!

It might be also that fundamental constants had not the same value in the early history of the Universe. 
%The right panel of Fig.\~ref{fig3} shows how the chemical composition of the core
%at the end of the core He-burning phase is modified when the binding energy of deuterium is changed.
%A change of the fine structure constant modifies the binding energy of deuterium.
%We see that variations of a few percents, which corresponds to variations of a few 10$^{-5}$ of the
%fine structre constant, implies very important changes of the core chemical composition at the end of the He-burning phase and thus deeply affect the advanced evolutionary stages of Pop III massive stars and their nucleosynthesis (Ekstr\"om et al. in preparation).
The chemical composition of the stellar cores at the end of the He-burning phase depends sensitively on the exact value of the fine structure constant, because variations of the fine structure constant implies changes of the 3$\alpha$ reaction rate. For instance, a variation
of a few 10$^{-5}$ suffices for instance to avoid any production of oxygen by core He-burning!
This deeply affects the advanced evolutionary stages of Pop III massive stars and of course their nucleosynthesis (Ekstr\"om et al. in preparation).

\section{Massive star evolution in the low metallicity regime}

At low (but non-zero) metallicities, rotational mixing plays a dominant role and produce two
important effects:
%This rotational mixing has a deep impact on the nucleosynthesis of some elements and on the way stars are losing mass.

{\it First, in this metallicity range, rotating models produce large amounts of primary nitrogen.}
Although rotational mixing in Pop III stars
is by far not a negligible effect, it remains at a relatively modest level due to the absence of
strong contraction at the end of the core H-burning phase. On the contrary, when $Z \ge \sim 10^{-10}$, the physical
conditions during the core H-burning phase and the core He-burning phase are so different that
a strong contraction occurs at the end of the core H-burning phase leading to strong mixing and
important primary nitrogen production\footnote{The presence of a very little amount of metals suffices to boost the efficiency of rotational mixing and the importance of mass loss. In that respect metallicity is like
the salt of the cosmos: a small amount is sufficient to enhance its flavor!}. 
For metallicities higher than about 0.001, rotational mixing is not efficient enough for triggering
important primary nitrogen production (at least for the rotational velocities corresponding
to the observed ones at this metallicity) and thus rotational mixing, although still
important for explaining the surface enrichments does no long change the stellar yields as much as at very low metallicity.
Thus primary nitrogen production does appear to go through a maximum when the metallicity decreases from $Z=0.001$ to $Z=0$, supposing that, at the different metallicities, the initial angular momentum content of a given initial mass star on the ZAMS remains more or less constant.

{\it Second, in this metallicity range, rotational mixing, by increasing the CNO surface abundances, might trigger important mass losses through radiatively driven stellar winds.} Indeed, 
another consequence from the strong contraction at the end of the core H-burning phase is 
a rapid evolution to the red in the HR diagram. Due to rotational mixing, the opacity in the outer layers increases with time, making appear 
an outer convective zone. This convective zone
dredges-up at the surface great quantities of CNO elements.
In our 60 M$_\odot$ stellar model with $Z=10^{-8}$ and $\upsilon_{\rm ini}=800$ km s$^{-1}$, the CNO
content at the surface amounts to one million times the one the star had at its birth (Meynet et al. 2006). Therefore the global metallicity at the surface becomes equivalent to that of a LMC stars while the star began its life with a metallicity which was about 600 000 times lower! If we apply the same rules used at higher metallicity relating the mass loss rate to the global metallicity, we obtain that the star may lose about half
of its initial mass due to this effect\footnote{Note that at the moment it is not possible to know if such a rule would apply in those circumstances, {\it i.e.} for typically a 60 M$_\odot$ Pop III star with an effective temperature of about 6000 K and a luminosity log$L/L_\odot=6.1$.}. 

There are at least four striking observational facts which might receive an explanation based on massive, metal poor, fast rotating stellar models.

\begin{figure}[t]
\begin{center}
\includegraphics[width=3in,height=2.7in]{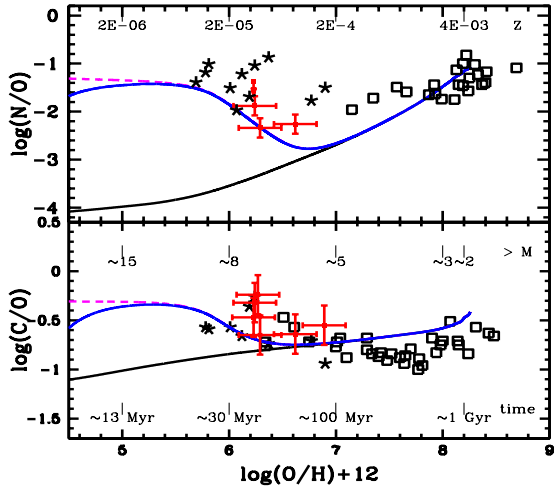}
%\hfill
%\includegraphics[width=1.5in,angle=-90]{MeynetFig2.ps}
\caption{Variation of the N/O and C/O ratios as a function the O/H ratios. N/O data points for halo stars are from Israelian et al.
 (2004, open squares ) and of Spite et al. (2005, stars). C/O points with error bars are for DLA systems from
 Pettini et al. (2008), C/O data points for halo stars are from Cayrel et al. (2004). The lower continuous curve is the chemical evolution model obtained with the stellar yields of slow rotating $Z=10^{-5}$ models from Meynet \& Maeder (2002) and Hirschi et al. (2004). The dashed line includes the yields of fast rotating $Z=10^{-8}$ models from Hirschi (2007) at very low metallicity. The intermediate curve is obtained using the yields of the $Z=0$ models presented in Ekstr\"om et al. (2008) up to $Z=10^{-10}$. The chemical evolution models are from
 Chiappini et al. (2006a).
%{\it Right panel}: s-process distributions between $^{57}$Fe and $^{138}$Ba normalized to solar for the 25 M$_\odot$ and [Fe/H]=-4 at the end of the convective C-burning shell. The horizontal line corresponds to the $^{16}$O overabundance in the C shell (thick line), multiplied and divided by two (thin lines). Isotopes of the same element are connected by a line. The cases presented are the following: {\it i)} non-rotating model (black triangles) and X($^{22}$Ne)$_{\rm ini}=5.21 \times 10^{-5}$; {\it ii)} rotating model (open squares, full squares for the s-only isotopes) and X($^{22}$Ne)$_{\rm ini}=5.0 \times 10^{-3}$; {\it iii)} rotating model (open circles, full circles for the s-only isotopes) and X($^{22}$Ne)$_{\rm ini}=1.0 \times 10^{-2}$. Figure from Pignatari et al. (2008).
}\label{figncs}
\end{center}
\end{figure}
\vskip 2mm
\noindent\underline{Normal metal poor halo stars}
\vskip 2mm
The recent observations of the surface abundances of very metal poor halo stars\footnote{These stars are in the field and present [Fe/H] as low as -4, thus well below the metallicities of the globular clusters.} show the need of a very efficient mechanism for the production of primary nitrogen (Chiappini et al. 2005). As explained in Chiappini et al. (2006a), a very nice way to explain this very efficient primary nitrogen production is to invoke fast rotating massive stars. Very interestingly, fast rotating massive stars help not only in explaining the behavior of the N/O ratio at low metallicity but also those of the C/O. In Fig.~\ref{figncs},
predictions for the evolution of N/O and C/O of chemical evolution models using different sets of yields are compared (Chiappini et al. 2006a\footnote{The details of the chemical evolution models can be found in Chiappini et al. (2006b), where they show that such a model reproduces nicely the metallicity distribution of the Galactic halo. This means that the timescale for the enrichment of the medium is well fitted.}). We see that the observed N/O ratio is much higher than what is predicted by a chemical evolution model using the yields of the slow-rotating $Z=10^{-5}$ models from Meynet \& Maeder (2002) down to $Z=0$. When adding the yields of the fast-rotating $Z=10^{-8}$ models from Hirschi (2007)
the fit is much improved. The same improvement is found for the C/O ratio, which presents an upturn at low metallicity. Thus these comparisons support fast rotating massive stars as the sources
of primary nitrogen in the galactic halo.
 
High N/O and the C/O upturn of the low-metallicity stars are also observed in low-metallicity DLAs  (Pettini et al. 2008, see the crosses in Fig.~\ref{figncs}). We note that the observed points are
below the points for the halo stars in the N/O versus O/H plane. This may be attributed to two causes: either the observed N/O ratios observed in halo stars are somewhat overestimated or
the difference is real and might be due to different star formation histories in the halo and in DLAs. Let us just discuss these two possibilities. 

Measures of nitrogen abundances at the surface of very metal poor stars is quite challenging, much more than the measure of nitrogen in the interstellar medium as is done for the DLAs, therefore one expects that the data for DLAs suffer much smaller uncertainties than those for halo stars. 
In that respect 
the observed N/O ratios
in DLAs give more accurate abundances than halo stars. 
%In case both DLAs and the galactic halo had the same star formation history, {\it i.e.} are the result of an intense and rapid star formation episode, then the chemical evolution models presented in Fig.~\ref{NOCO} can also apply to DLAs. 
%We see that the DLA data, although not extending as far in the low Z region and being slightly below than the halo data, 
%still require models with fast rotation to be fitted.
Most probably the star formation history in DLAs is not the same as in the halo. While, as recalled above, in the halo we see the result of a strong and rapid star formation episode, in DLAs one might see the result of  much slower and weaker star formation episodes. In that case, both massive stars and intermediate mass stars contributed to the build up of the chemical abundances and the chemical evolution models presented in Fig.~\ref{figncs} do no longer apply to these systems (see Chiappini et al. 2003; Dessauges-Zavadsky 2007 for chemical evolution models of DLAs). It will be very interesting to study the results of chemical evolution models adapted to this situation and accounting for stellar yields from both rotating massive and intermediate stars. Let us just mention at this stage that primary nitrogen production in metal poor intermediate mass stars is also strongly
favored when rotational mixing is accounted for (Meynet \& Maeder 2002). Thus also in that case, rotation may play a key role.

The primary nitrogen production is accompanied by other interesting features such as the production
of primary $^{13}$C (see Chiappini et al. 2008).
Production of primary $^{13}$C by massive stars can explain the low $^{12}$C/$^{13}$C ratios found recently by Spite et al. (2006) for normal very metal poor halo stars without invoking the contribution of AGBs which, according to the chemical evolution models for the halo, did not play a major role bellow [Fe/H]$\sim$ -2.5. This is important as  recent observational results (Melendez \& Cohen 2007) found  the $^{25}$Mg/$^{24}$Mg, $^{26}$Mg/$^{24}$Mg ratios in halo metal poor stars to be low, again suggesting that AGB stars would have played a minor role below [Fe/H] $\sim$ -2.0.
The primary nitrogen production is also accompanied by the production 
of primary $^{22}$Ne. Primary $^{22}$Ne is produced by diffusion of primary nitrogen from the H-burning shell to the core He-burning zone, or by the engulfment of part of the H-burning shell by the growing He-burning core. These processes
occur in rotating massive star models (Meynet \& Maeder 2002; Hirschi 2007). In the He-burning zone, $^{14}$N
is transformed into $^{22}$Ne through the classical reaction chain
$^{14}$N($\alpha$,$\gamma$)$^{18}$F($\beta^+$ $\nu$)$^{18}$O($\alpha$,$\gamma$)$^{22}$Ne.

In the He-burning zones (either in the core at the end of the core He-burning phase or in the
He-burning shell during the
core C-burning phase and in the following convective C-burning shell), neutrons are released through the reaction
$^{22}$Ne($\alpha$,n)$^{25}$Mg. 
These neutrons then can either be captured by iron seeds and produce s-process elements or be captured by light neutron poisons and
thus be removed from the flux of neutrons.
The final outputs of s-process elements will
depend on at least three factors: the amounts of 1.- $^{22}$Ne, 2.- neutron poisons and 3.- iron seeds. In standard models (without rotation), when the metallicity decreases, the amount of $^{22}$Ne decreases
(less neutrons produced), the strength of primary neutron poisons becomes relatively more important in particular for [Fe/H]$\le$-2 with respect to solar, and the amount of iron seeds also decreases (e.g., Raiteri et al. 1992).
Thus very small quantities of s-process elements are expected. 
%(see the triangles in the right panel of  Fig.~\ref{figncs}). 
When primary nitrogen and therefore primary $^{22}$Ne is present in quantities as given by rotating models
which can reproduce the observed trends for the N/O and C/O ratios in the
halo stars, then a very different output is obtained (Pignatari et al. 2008). 
The abundances of several s-process elements are increased by many orders of magnitudes. In particular, the elements are produced in the greatest quantities in the atomic mass region between strontium and barium,
and no long in the atomic mass region between iron and strontium as in the case of standard models.

These first results show that some heavy s-process elements, not produced in standard models (without rotation), might be produced
in significant quantities in metal poor rotating stellar models. It will be very interesting in the future to find some non ambiguous signature of the occurrence of this process in the 
abundance pattern of very metal poor halo stars.
\vskip 2mm
\noindent\underline{C-rich Extremely Metal Poor Stars (CEMP)}
\vskip 2mm
Below about [Fe/H] $<$ -2.5, a significant fraction of very iron-poor stars are C-rich
(see the review by Beers \& Christlieb (2005). Some of these stars show no evidence of $s$-process enrichments by AGB stars and are thus likely formed from the ejecta of massive stars. The problem is how to explain the very high abundances with respect to iron of CNO elements.
As shown by Meynet et al. (2006) and Hirschi (2007) the matter released
in the winds of fast rotating very metal poor massive stars is enriched in both H- and He-burning products and present striking similarities with the abundance patterns observed at the surface of CEMPs and these authors proposed that at least some of these stars might be formed from the winds of such objects. Note that stars formed in that way should also be He-rich!
It is likely that rotation also affects the composition of the ejecta of intermediate mass stars. Meynet et al. (2006) predict the chemical composition of the envelope of a 7 M$_\odot$ E-AGB star which have been enriched by rotational mixing. The composition also presents striking similarities with the abundance patterns observed at the surface of CEMPs. The presence of overabundances of fluorine (Schuler et al. 2007) and of $s$-process elements might be used to discriminate between massive and intermediate mass stars.
\vskip 2mm
\noindent\underline{He-rich stars in globular clusters}
\vskip 2mm
Indirect observations indicate the presence of very helium-rich stars in the globular cluster $\omega$Cen (Piotto et al. 2005). Stars with a mass fraction of helium, $Y$,  equal to 0.4 seem to exist, together with a population of normal helium stars with $Y=0.25$. 
Other globular clusters appear to host helium-rich stars (Caloi \& D'Antona 2007), thus
the case of $\omega$Cen is the most spectacular but not the only one.
There is no way for these very low mass stars to enrich their surface in such large amounts of helium and one possibility is that they have acquired their He abundance from 
the protostellar cloud from which they formed (an alternative would be through mass transfer in close binary systems).
Where does this helium come from? We proposed that it was shed away by the winds of metal poor fast rotating stars (Maeder \& Meynet 2006). 
\vskip 2mm
\noindent\underline{Chemical anomalies in globular clusters}
\vskip 2mm
In globular clusters, stars made of material only enriched in H-burning products have been observed (see the review by Gratton et al. 2004). Probably these stars are also enriched in helium and thus this observation is related to the one reported just above. The difference is that proper abundance studies can be performed for carbon, nitrogen, oxygen, sodium, magnesium, lithium, fluorine \dots, while for helium only indirect inferences based on the photometry can be made. 
Decressin et al. (2007a)  propose that the matter from which the stars rich in H-burning products are formed, has been released by slow winds of fast rotating massive stars. Other authors have proposed AGB stars as the main supplier of the material from which the Na-rich and O-poor stars are formed (see {\it e.g.} D'Antona \& Ventura 2008).
The massive star origin presents however some advantages: first a massive star can induce star formation in its surrounding, thus two effects, the enrichment and the star formation can be triggered by the same cause. Second, the massive star scenario allows to use a less flat IMF than the scenario invoking AGB stars (Prantzos \& Charbonnel 2006). The slope of the IMF might be even a Salpeter's one in case the globular cluster lost a great part of its first generation stars by tidal stripping (Decressin et al. 2007b; Decressin et al. 2008).

All the above observations seem to point toward the same direction, an important population of spinstars at low Z. Of course alternative explanations exist for all these features. One advantage
of those presented above is that they rely on one unique physical process: rotational mixing!

\section{Massive star evolution in the near solar metallicity regime}

%The near solar metallicity regime is the domain where detailed comparisons between stellar models and observations can be performed. This is thus the key metallicity range where models can be checked
%and constrained.

In the near solar metallicity regime, rotation and mass loss by stellar winds  are of
similar importance. Neither of the two aspects can be neglected. This metallicity range
is also the one in which models can be checked and calibrated by comparisons with well observed features either of individual stars or of stellar populations. 
Among these observed features let us cite
\begin{itemize}
\item The observed changes of the surface abundances.
\item The observed changes of the surface velocities.
\item The shape of fast rotating stars measured by interferometric technics, variation of the effective temperature with the colatitude, measures of wind anisotropies, shape
of nebulosities resulting from outbursts.
\item The sizes of the convective cores as deduced from  asterosismic analysis.
Asterosismic observations can also constrain the interior variation 
of the angular velocity.
\item The width of the Main Sequence band.
\item The existence and variation with $Z$ of the populations of Be stars.
\item The variation with $Z$ of the blue to red supergiant ratio.
\item The variation with $Z$ of the Wolf-Rayet populations.
\item The rotation rates of young pulsars.
\item The variation with $Z$ of different core collapse supernova types.
\end{itemize}
As a general statement, it does appear that models including the effects of rotation provide a much better fit to most of the above observed features. Here we shall discuss two of them, the cases of
surface abundances and the variation with $Z$ of the ratio of type Ibc to type II supernovae. 

A rotating star is predicted to present some nitrogen surface enrichment already during the main sequence. The amplitude of the nitrogen enrichment at the surface depends on the initial mass (increases with the mass), the age (increases with the age) and the initial rotational velocity
(increase with the initial velocity).
Thus we see that the nitrogen surface abundance is at least a function of three parameters:
mass, age and velocity. 
This is correct as long as we consider stars with a given initial composition (rotational mixing is more efficient at low Z) and whose evolution is not affected by a close binary companion.

\begin{figure}[t]
\begin{center}
\includegraphics[width=2.7in,height=2.3in]{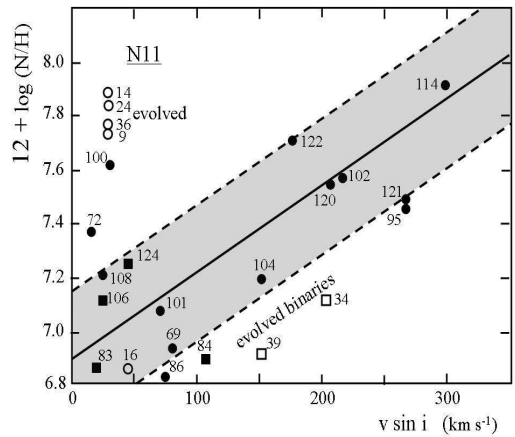}
\caption{The N abundance (in a scale where $\log H=12.0$) 
as a function of $v \sin i$ for the MS stars (black dots) in N11 with masses between
14 and 20 M$_{\odot}$ according to Hunter (2008, private communication). The binaries are shown by a square. The evolved stars
in a band of 0.1 dex in $\log T_{\mathrm{eff}}$ beyond the end of the MS are shown with open symbols.
The gray band indicates uncertainties of $\pm 0.25$ dex. Figure taken from Maeder et al. (2008).}
\label{NHN11}
\end{center}
\end{figure}

To see a relation between N-enrichment and velocity, it is necessary
to use stars with different rotational velocities but having similar masses and ages. 
In Fig.~\ref{NHN11}, such a relation is shown for stars in the N11 SMC cluster (Maeder et al. 2008), where
the sample is limited to the stars in the mass range 14 to 20 M$_{\odot}$ on the basis of  the data provided by Hunter (2008, private communication) and in the formal MS band  as given by Fig.~34 from Hunter et al. (2007).

We see that the bulk of stars in N11 shows a relation of the excess of N/H depending on 
$v \sin i$ (the mean square root of the data for the MS band stars is 0.23 dex from the data by Hunter 2008, private communication, the scatter in $v \sin i$ is not given). The  amplitude of  the (N/H) is about 0.6 dex for velocities of  200 km s$^{-1}$, slightly higher than the value obtained from
rotating stellar models for $Z=0.02$ for the corresponding masses (see Fig.~1 in Maeder et al. 2008).

When data samples, limited in masses and ages, are used, a very nice correlation is found between
the surface N-enrichment and $\upsilon\sin i$ (see also Fig. 4 in Maeder et al. 2008), 
supporting a N enrichment depending on rotational velocities. Stars beyond the end of the MS phase
do not obey to  such a relation, because their velocities 
converge toward low values (see Fig.~12 by Meynet \& Maeder 2000). 
A fraction, which we estimate to be   $\sim20$ \%
of the stars, may escape from the relation as a result of binary evolution, either by tidal mixing or mass transfer.

Core collapse supernovae of type Ib and Ic are very interesting events for many reasons. One of them is that in four cases, the typical spectrum of a type Ic supernova has been observed together with a long soft Gamma Ray Burst (GRB) event (Woosley \& Bloom 2006). Also, recent observations 
(Prieto et al. 2008) present new values for the variation with the metallicity of the number ratio 
(SN Ib+SN Ic)/SN II to which theoretical predictions can be compared. Finally, according at least to single star scenarios, these supernovae arise from the most massive stars. They offer thus a unique opportunity to study the final stages of these objects
which have a deep impact on the photometric and spectroscopic evolution of galaxies and also contribute to its chemical evolution .

We shall now discuss the predictions of single star models for the type Ib/Ic supernovae frequency.
Since these supernovae do not show any H-lines in their spectrum, they should have as progenitors stars having removed {\it at least} their H-rich envelope by stellar winds, {\it i.e.} their progenitors should be WR stars of the WNE type (stars with no H at their surface and presenting He and N lines) or of the WC/WO type (stars with strong overabundances of He-burning products at their
surface, mainly carbon and oxygen).

\begin{figure}[t]
\includegraphics[width=2.3in,height=2.3in]{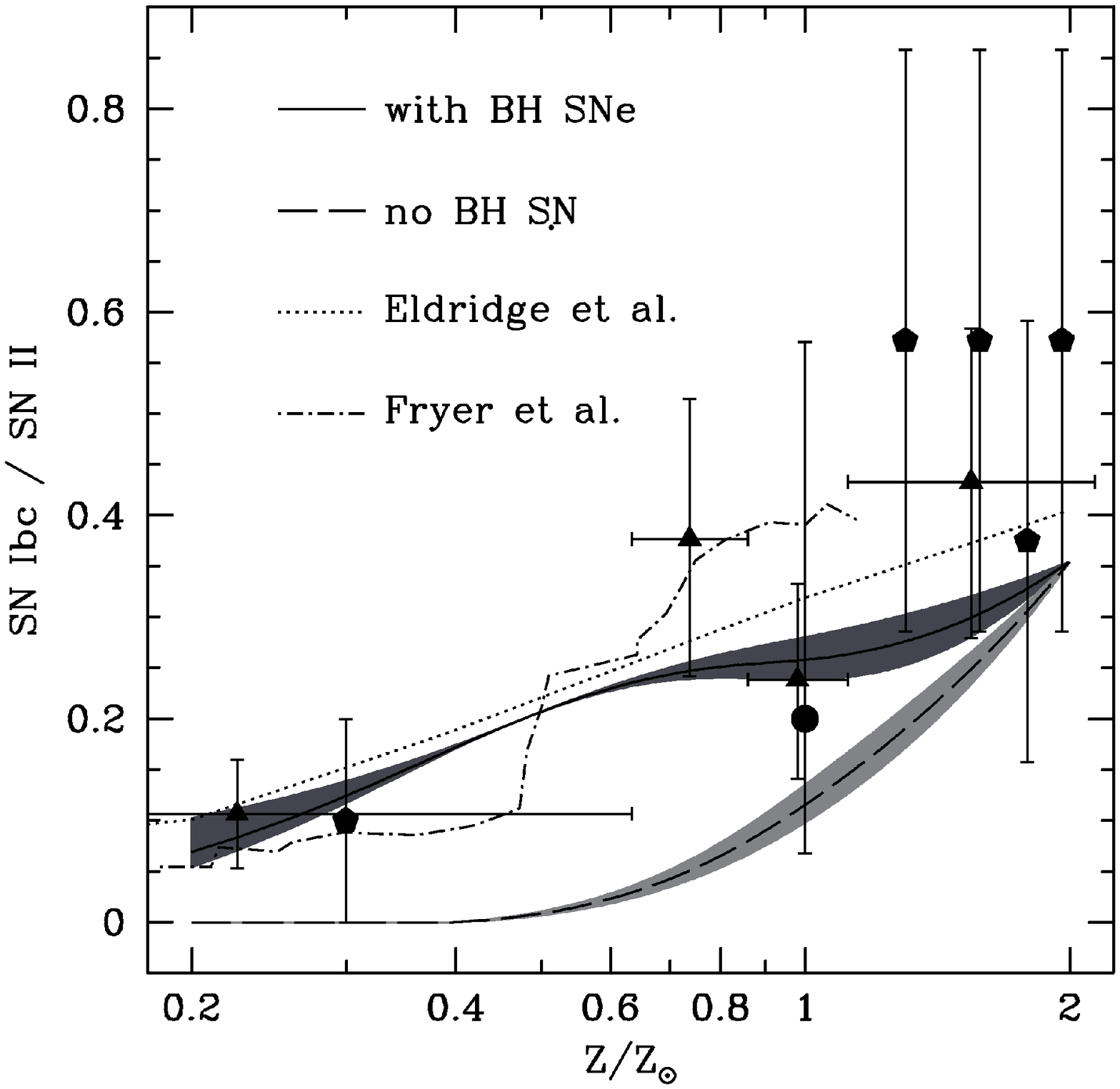}
\hfill
\includegraphics[width=2.3in,height=2.3in]{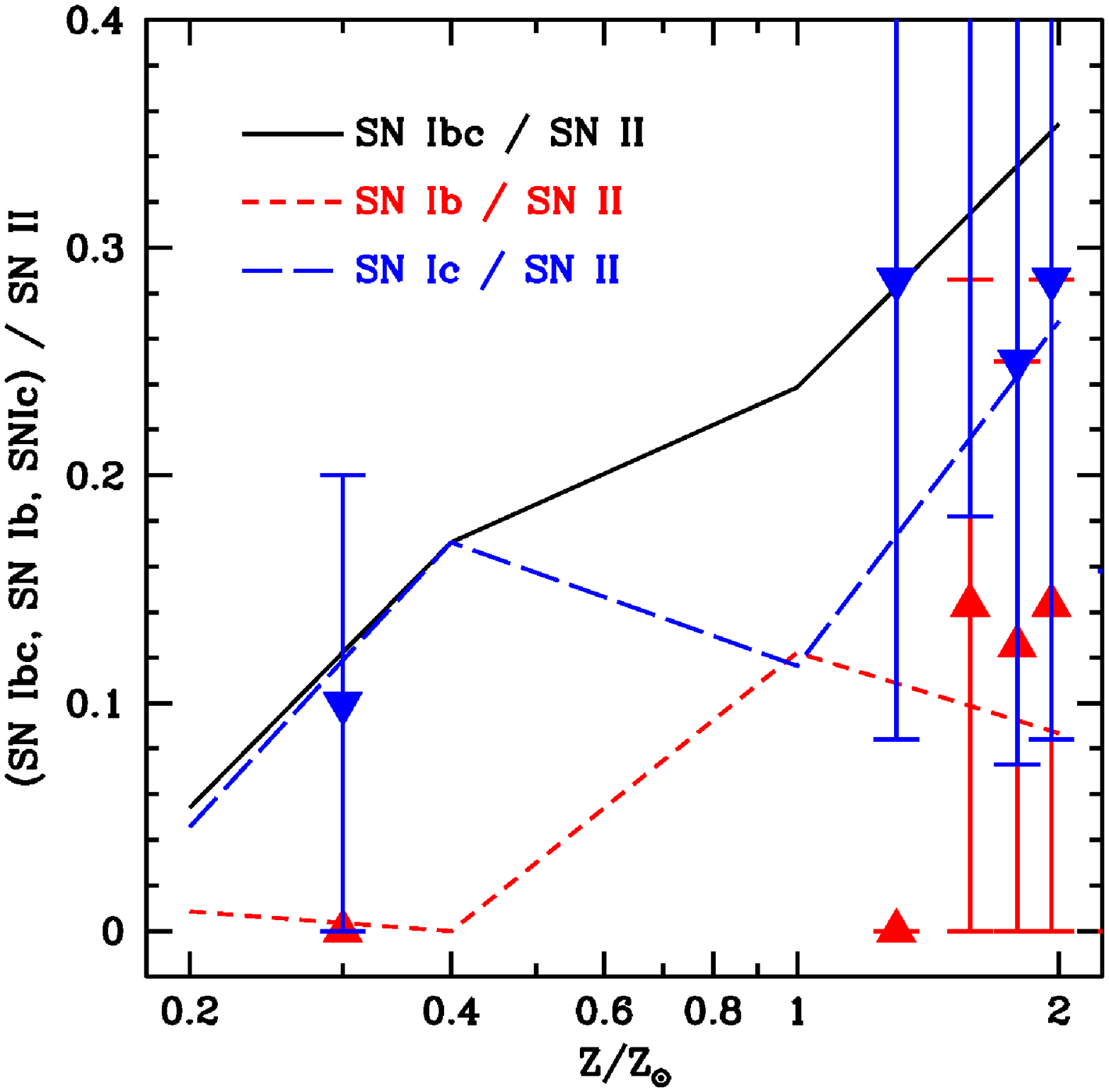}
\caption{{\it Left panels}: Rate of SN Ibc / SN II if all models produce a SN (solid line) or if models producing a black holes do not explode in a SN (dashed line). Grey areas are the corresponding estimated errors from our models. Pentagons are observational data from Prieto et al. (2008), triangles are data from Prantzos \& Boissier (2003) and circle is the measurement at solar metallicity from Cappellaro \& Turatto (2001). The dotted line represents the binary models of Eldridge et al. (2008), and the dotted-dashed line the rate obtained with the binary models of Fryer et al. (2007).
{\it Right panel}: Rates of SN Ic / SN II (blue long--dashed line), SN Ib / SN II (red short--dashed line) and SN Ibc / SN II (black solid line). The points are extracted from the data of Prieto et al. (2008): triangles (red) represent the observed SN Ib / SN II rate, and upside--down triangles (blue) the observed SN Ic / SN II rate. Each triangle corresponds to a sample of 11 core collapse SNe (color figure available online). Figure taken from Georgy et al. submitted.}\label{RateBlackHole}
\end{figure}

Considering that all models ending their lifetime as a WNE or WC/WO phase will explode as a type Ibc supernova, it is possible to compute the variation with the metallicity of the number ratio of type Ibc to type II supernovae. The result is shown in Fig.~\ref{RateBlackHole} (left panel, see also Meynet \& Maeder 2005). One sees that
this ratio increases with the metallicity. This is due to the fact that at higher metallicity, the minimum initial mass of stars ending their life as WNE, or WC/WO stars is lower than at lower
metallicities. Single star models can reasonably well reproduce the observed trend with the metallicity. They however give slightly too small values with respect to the observations, which may indicate that a portion of the type Ibc supernovae may originate from close binary evolutions. 
Models accounting
for single and binary channel (but without rotation) are shown as a dotted line (Eldridge et al. 2008). They provide a good
fit to the observations. But in that case most of the supernovae originate from the binary channel,
leaving little place for the single star scenario. These models would also predict that most of the
WR stars are the outcome of close binary evolution. This does not appear to be confirmed by the
observations of Foellmi et al (2003ab).
Most likely, both the single and binary channel contribute.

The right panel of Fig.~\ref{RateBlackHole} shows how the number ratio of type Ib and type Ic supernovae considered separately compare with the observations. 
%To link the  results with the type of the supernova event we adopted the following rules:
%as long as some hydrogen is present in the ejecta, a type II supernova will occur. 
%We considered that all supernovae ejecting less than about $0.6 \,\mathrm{M}_\odot$ of helium are of type Ic. All progenitors satisfying neither the criterion for becoming a type II (no H present
%in the ejecta), neither the one for becoming a type Ib (more than $\sim$0.6 M$_\odot$ of He) are %considered to give rise to type Ib SN events. 
We see that a good agreement is obtained although the observations are still scarce and based on a small number of cases.

Of course the situation may change in case, when a Black Hole (BH) is formed, no SN event occurs.
We computed new (SN Ib + SN Ic)/ SN II ratios 
with the assumption that all models massive enough to form a black hole do not produce a SN. 
Comparing with the observed rates in the left panel of Fig.~\ref{RateBlackHole} (see dashed line)
we see that in the case no supernova event occurs when a BH is formed, 
single star models might still account for a significant fraction of the type Ibc supernovae
for $Z > 0.02$.
At $Z=0.004$ all type Ibc should arise from other evolutionary scenarios. A possibility would be
in that case to invoke close binary evolution with mass transfer. 

Probably, the hypothesis according to which no supernova event is associated when a BH is formed, is too restrictive. For instance, 
the collapsar scenario for Gamma Ray Bursts (Woosley 1993)  needs the formation of a black holes (Dessart, private communication) and this formation is at least accompanied in some cases by a type Ic supernova event. 
Also the observation of 
the binary system GRO J1655-40 containing a black hole (Israelian et al. 1999) suggest that a few stellar masses have been ejected  and that a SN event occurred when the BH formed. This is deduced from the important chemical anomalies observed at the surface of the visible companion, chemical anomalies whose origin is attributed to the fact that the (now visible) companion accreted part of the SN ejecta. This gives some
support to the view that, at least in some cases, the collapse to a BH does not prevent mass ejection and a supernova event to occur.

As emphasized at the beginning of this section, rotation and mass loss probably are both important
in this metallicity range. Impact of these two processes on various outputs of solar metallicity stellar models can be found in Heger \& Langer (2000), Meynet \& Maeder (2000; 2003), Hirschi et al. (2005a).

\section{Massive star evolution at metallicities above the solar metallicity}

Above solar metallicities, radiative line driven winds become the dominant factor affecting the evolution of massive stars. In models with moderate rotation, we note however that the effects of rotational mixing are still important but their impact is less apparent being somewhat mixed with those of the stellar winds. From a theoretical point of view we can note two features which do appear different at high metallicity: first the chemical enrichment of the interstellar medium is different. Indeed, as been shown by Maeder (1992), when the stellar winds are important, greater quantities of 
products from the early phases of the evolution of stars are ejected into the interstellar medium
when comparison is made with the chemical abundances of the ejecta of a similar star which would have released its outer layers only at the supernova stage. Physically this comes from the fact that when a layer of stellar material is ejected by stellar winds, it is ejected at an early phase of the evolution of the star, when the layer has only been partially processed by the nuclear reactions. This
allows the production of some elements to be enhanced and other to be decreased with respect to a star releasing its
outer layers only at the supernova stage.
Computations shows that at high metallicity, 
greater quantities of helium, carbon (Maeder 1992), fluorine (Meynet \& Arnould 2000; Palacios et al. 2005a), aluminum 26 (Palacios et al. 2005b), s-process elements (Arnould et al. 2006) will be ejected.
This may have interesting consequences for the chemical evolution of the galaxies at high metallicity.

%A star which loses a lot of material by stellar winds may differently enrich the interstellar medium
%in new elements, compared to a star which would have retained its mass all along until the supernova explosion. As Maeder (1992) pointed out, when the stellar winds are strong, material partially processed
%by the nuclear reactions will be released, favoring some species (as helium and carbon which would be otherwise partially destroyed if
%remained locked into the star) and disfavoring other ones (as e.g. oxygen which would be produced by further transformation of the species which are wind-ejected).

Another interesting difference which occurs at high metallicity is the evolution of the angular momentum. As we explained in Sec.~2, one expects that at high metallicity, angular momentum is more easily transported from the core to the surface and more easily ejected at the surface by the stellar winds. Thus, everything being equal, one would expect that the angular momentum of the central regions will be lower when the metallicity is higher. Together with the fact that Black-Hole formation is probably more difficult at high $Z$ (also because of the strong stellar winds), this makes the formation of collapsars which are considered as a serious candidate for long soft gamma ray burst much less favorable at high metallicity (see e.g. Hirschi et al. 2005b). In case the rotation rate of young pulsar depends
to some extent to the rotation rate of the core at the presupernova stage, the above line of reasoning would lead to the conclusion that the rotation rate of young pulsars should be slower
in metal rich regions than in metal poor ones.

\section{Conclusion}

Massive star evolution is at the crossroad of many topical astrophysical problems: their link with $gamma$-ray line astrophysics, the origin of galactic cosmic rays, that of isotopic anomalies in the meteorites, the
many puzzling observed features related to very metal poor stars in the field of the halo, to the stars in globular clusters, to the progenitors of core collapse supernovae and gamma ray bursts, to neutron stars and black holes, place them at the heart of modern astrophysics.  Moreover they represent unique tool to probe the distant universe and constitute important sources of radiations, of new synthetised nuclei  and of momentum in galaxies.
Still major improvements of massive star models are needed. Great deal of efforts have been
made to provide more realistic models accounting for the effects of 
mass loss and rotation. Aspects
as the effects of accretion processes during the pre-main sequence phase, magnetic fields, tidal forces in close binaries, still remain to be further explored and represent wonderful challenges for future works.

\subsection*{References}

{\small

\bref
Arnould, M., Goriely, S., Meynet, G. 2006, A\&A 453, 653

\bref
Beers, T.\,C., Christlieb, N. 2005, ARAA 43, 531

\bref
Caloi, V., D'Antona, F. 2007, A\&A 463, 949

\bref
Cappellaro, E., Turatto, M. 2001, The influence of binaries on stellar population studies, Dordrecht: Kluwer Academic Publishers, 2001, xix, 582 p. Astrophysics and space science library (ASSL), Vol. 264. ISBN 0792371046, p.199

\bref
Casagrande, L., Flynn, C., Portinari, L., Girardi, L., Jimenez, R. 2007, MNRAS 382, 1516

\bref
Cayrel, R., Depagne, E., Spite, M. 2004, A\&A 416, 1117

\bref
Chiappini, C., Matteucci, F., Meynet, G. 2003, A\&A 410, 257

\bref
Chiappini, C., Matteucci, F., Ballero, S.\,K. 2005, A\&A 437, 429

\bref
Chiappini, C., Hirschi, R., Meynet, G., Ekstr\"om, S., Maeder, A., Matteucci, F. 2006a,
     A\&A Letters 449, 27
     
\bref
Chiappini, C., Hirschi, R.,  Matteucci, F., Meynet, G., Ekstr\"om, S., Maeder, A. 2006b,
     in ``Nuclei in the Cosmos IX'', Proceedings of Science, 9 pages (arXiv:astro-ph/0609410)     
     
\bref
Chiappini, C., Ekstr\"om, S.,Hirschi, R., Meynet, G.,  Maeder, A., Charbonnel, C. 2008,
     A\&A Letters 479, 9 

\bref     
Cayrel, R., Depagne, E., Spite, M. et al. 2004, A\&A 416, 1117   

\bref
D'Antona, F., Ventura, P. 2008, The Messenger 134, 18 

\bref
Decressin, T., Baumgardt, H.,  Kroupa, P. 2008, A\&A 492, 101

\bref
Decressin, T., Charbonnel, C., Meynet, G. 2007b, A\&A 475, 859

\bref
Decressin, T., Meynet, G., Charbonnel, C., Prantzos, N., Ekstr{\"o}m, S.
2007a, A\&A 464, 1029

\bref
Dessauges-Zavadsky, M., Calura, F., Prochaska, J.\,X., D'Odorico, S., Matteucci, F.
2007, A\&A 470, 431

\bref
Domiciano de Souza, A., Kervella, P., Jankov, S., Abe, L., Vakili, F., di Folco, E., Paresce, F.
2003,  A\&A 407, L47

\bref
Dwarkadas, V.\,V., Owocki, S.\,P. 2002, ApJ 581, 1337

\bref 
Eddington, A.\,S. 1926, in ``The Internal Constitution of The Stars'',
Cambridge University, Cambridge.

\bref
Ekstr\"om, S., Meynet, G., Chiappini, C., Hirschi, R., Maeder, A. 2008a, A\&A 489, 685

\bref
Ekstr\"om, S., Meynet, G., Maeder, A. 2008b, IAU Symposium 250, p. 209

\bref
Eldridge, J.\,J., Izzard, R.\,G., Tout, C.\,A. 2008, MNRAS 384, 1109

\bref
Foellmi, C., Moffat, A.\,F.\,J., Guerrero, M.\,A. 2003a, MNRAS 338, 360

\bref
Foellmi, C., Moffat, A.\,F.\,J., Guerrero, M.\,A. 2003b, MNRAS 338, 1025

\bref
Fryer, C.\,L., Mazzali, P.\,A., Prochaska, J. et al. 2007, PASP 119, 1211

\bref
Gratton, R., Sneden, C., Carretta, E. 2004, ARA\&A 42, 385 

\bref
Greif, T.\,H., Bromm, V. 2006, MNRAS 373, 128

\bref
Heger, A., Langer, N. 2000, ApJ 544, 1016

\bref
Heger, A., Fryer, C.\,L., Woosley, S.\,E., Langer, N., Hartmann, D.\,H. 2003, ApJ 591, 288

\bref
Hirschi, R. 2007, A\&A 461, 571

\bref
Hirschi, R., Meynet, G., Maeder, A. 2004, A\&A 425, 649

\bref
Hirschi, R., Meynet, G., Maeder, A. 2005a, A\&A 433, 1013

\bref
Hirschi, R., Meynet, G., Maeder, A. 2005b, A\&A 443, 581

\bref
Hunter, I., Dufton, P.\,L., Smartt, S.\,J. et al. 2007, ApJ 466, 277

\bref
Iocco, F., Bressan, A., Ripamonti, E., Schneider, R., Ferrara, A., Marigo, P. 2008,
MNRAS 390, 1655

\bref
Israelian, G. et al. 1999, Nature 401, 142

\bref
Israelian, G., Ecuvillon, A., Rebolo, R., García-López, R., Bonifacio, P., Molaro, P.
2004, A\&A 421, 649

\bref
Johnson,J.\,L., Bromm, V. 2006, MNRAS 366, 247

\bref
Kudritzki, R.\,P., Pauldrach, A.\,W.\,A, Puls, J. 1987, A\&A 173, 293

\bref
Lamers, H.J.G.L.M., Snow, T.P., Lindholm, D.M. 1995, ApJ 455, 269       

\bref
Leitherer, C., Robert, C., Drissen, L. 1992, ApJ 401, 596

\bref
Maeder, A. 1987, A\&A 158, 179

\bref
Maeder, A. 1992, A\&A 264, 105

\bref
Maeder, A. 1997, A\&A 321, 134 

\bref
Maeder, A. 1999, A\&A 347, 185

\bref
Maeder, A. 2003, A\&A 399, 263

\bref
Maeder, A., Desjacques, V. 2001, A\&A 372, L9

\bref
Maeder, A., Meynet, G. 2000a, ARA\&A 38, 143

\bref
Maeder, A., Meynet, G. 2000b, A\&A 361, 159

\bref
Maeder, A., Meynet, G. 2001, A\&A 373, 555 (paper VII)

\bref
Maeder, A., Meynet, G. 2005, A\&A 440, 1041 

\bref
Maeder, A., Meynet, G. 2006, A\&A 448, L37

\bref
Maeder, A., Zahn, J.P. 1998, A\&A 334, 1000

\bref
Maeder, A., Grebel, E.K., Mermilliod, J.-C. 1999, A\&A 346, 459 

\bref
Maeder, A., Meynet, G., Ekstrom, S., Georgy, C. 2008, Comm. in Asteroseismology, Contribution to the Proceedings of the 38th LIAC, HELAS-ESTA, BAG, in press (arXiv:0810.0657)

\bref
Marigo, P., Chiosi, C., Kudritzki, R.-P. 2003, A\&A 399, 617

\bref
Martayan, C., Fr\'emat, Y., Hubert, A.-M. et al. 2006, A\&A 452, 273

\bref
Martayan, C., Fr\'emat, Y., Hubert, A.-M., Floquet, M., Zorec, J., Neiner, C.
2007, A\&A 462, 683

\bref
Mathis, S., Palacios, A., Zahn, J.-P. 2004, A\&A 425, 243

\bref
Matteucci, F. 2001, The chemical evolution of the Galaxy, Astrophysics and space science library, Volume 253, Dordrecht: Kluwer Academic Publishers, ISBN 0-7923-6552-6, 2001, XII + 293 pp.

\bref
Melendez, J., Cohen, J.G. 2007, ApJ 659, L25

\bref
Meynet, G., Arnould, M. 2000, A\&A 355, 176

\bref
Meynet, G.; Maeder, A. 2000, A\&A 361, 101

\bref
Meynet, G., Maeder, A.,  2002a, A\&A 381, L25 

\bref
Meynet, G., Maeder, A.,  2002b, A\&A 390, 561 

\bref
Meynet, G., Maeder, A.,  2003, A\&A 404, 975    

\bref
Meynet, G., Maeder, A.,  2005, A\&A 429, 581 

\bref
Meynet, G., Maeder, A. 2006, in Stars with the B[e] Phenomenon. ASP Conf. Ser., eds.
M. Kraus \& A.S. Miroshnichenko, 355, p. 27

\bref     
Meynet, G., Ekstroem, S., Maeder, A. 2006, A\&A 447, 623   

\bref
Mowlavi, N., Meynet, G., Maeder, A., Schaerer, D., Charbonnel, C. 1998, A\&A 335, 573

\bref
Owocki, S.P., Cranmer, S.R., Gayley, K.G. 1996, ApJ 472, L115

\bref
Palacios, A., Arnould, M., Meynet, G. 2005a, A\&A 443, 243

\bref
Palacios, A., Meynet, G., Vuissoz, C. 2005b, A\&A 429, 613

\bref
Pettini, M., Zych, B. J., Steidel, C. C., Chaffee, F. H. 2008, MNRAS 385, 2011 

\bref
Pignatari, M., Gallino, R., Meynet, G., Hirschi, R., Herwig, F., Wiescher, M. 2008, ApJ 687, L95

\bref
Piotto, G., Villanova, S., Bedin, L.\,R. 2005, ApJ 621, 777

\bref
Prantzos, N., Boissier, S. 2003, A\&A 406, 259

\bref
Prantzos, N., Charbonnel, C. 2006, A\&A 458, 135

\bref
Prieto, J.\,L., Stanek, K.\,Z., Beacom, J.\,F. 2008, ApJ 673, 999

\bref
Raiteri, C.\,M., Gallino, R., Busso, M. 1992, ApJ 387, 263

\bref
Schneider, R., Salvaterra, R., Ferrara, A., Ciardi, B. 2006, MNRAS 369, 825

\bref
Schuler, S.\,C., Cunha, K., Smith, V.\,V., Sivarani, T., Beers, T.\,C.; Lee, Y.\,S. 2007, ApJ 667, L81

\bref
Spite, M., Cayrel, R., Plez, B. 2005, A\&A 430, 655

\bref
Spite, M., Cayrel, R., Hill, V. et al. 2006, A\&A 455, 291

\bref
Spolyar, D., Freese, K., Gondolo, P. 2008, Physical Review Letters 100, 051101

\bref
Spruit, H.\,C. 2002, A\&A 381, 923

\bref
Talon, S. 2008, EAS Publications Series 32, 2008, 81

\bref
Talon, S., Zahn, J.P. 1997, A\&A 317, 749

\bref
Taoso, M., Bertone, G., Meynet, G., Ekstr\"om, S. 2008, Physical Review D 78, 123510

\bref
van Loon, J.Th., 2006, in \emph{Stellar Evolution at Low Metallicity: Mass Loss, Explosions, Cosmology}, H.J.G.L.M.~Lamers, N.~Langer, T.~Nugis, K.~Annuk, ASP Conf. Series 353, p. 211

\bref
Venn, K.\,A. 1999, ApJ 518, 405

\bref
Venn, K.\,A. Lambert, D.\,L. 2008, ApJ 677, 572

\bref
Venn, K.\,A., Przybilla, N. 2003, in CNO in the Universe, eds. Charbonnel, Schaerer \& Meynet, ASP Conf Ser. 304, p.20

%\bref
%Vink, J.\,S., de Koter, A. 2005, A\&A 442, 587

%\bref
%Vink, J.\,S., de Koter, A., \& Lamers, H.\,J.\,G.\,L.\,M. 2000, A\&A 362, 295

\bref
Vink, J.\,S., de Koter, A., \& Lamers, H.\,J.\,G.\,L.\,M. 2001, A\&A 369, 574

\bref
von Zeipel, H. 1924, MNRAS 84, 665 

\bref
Wisniewski, J.\,P., Bjorkman, K.\,S. 2006, ApJ 652, 458

\bref
Woosley, S.\,E. 1993, ApJ 405, 273

\bref
Woosley, S.\,E., Bloom, J.\,S. 2006, ARA\&A 44, 507

\bref
Yoon, S.-C., Iocco, F., Akiyama, S. 2008, ApJ 688, L1

\bref
Zahn, J.-P. 1992, A\&A 265, 115

}

\vfill

\end{document}